\documentclass[numberedappendix]{emulateapj}

\shorttitle{Short-Period Variables in Cetus and Tucana}
\shortauthors{Bernard et al.}

\begin{document}

\title{The ACS LCID Project. I. Short-Period Variables \\ in the
       Isolated Dwarf Spheroidal Galaxies Cetus \& Tucana\altaffilmark{1}}

\author{Edouard J. Bernard,\altaffilmark{2}
    Matteo Monelli,\altaffilmark{2}
    Carme Gallart,\altaffilmark{2}
    Igor Drozdovsky,\altaffilmark{2,3}
    Peter B. Stetson,\altaffilmark{4} \\
    Antonio Aparicio,\altaffilmark{2,5}
    Santi Cassisi,\altaffilmark{6}
    Lucio Mayer,\altaffilmark{7,8}
    Andrew A. Cole,\altaffilmark{9}
    Sebastian L. Hidalgo,\altaffilmark{2,10} \\
    Evan D. Skillman,\altaffilmark{10}
    Eline Tolstoy\altaffilmark{11}}

\altaffiltext{1}{Based on observations made with the NASA/ESA {\it Hubble Space
    Telescope}, obtained at the Space Telescope Science Institute, which is
    operated by the Association of Universities for Research in Astronomy,
    Inc., under NASA contract NAS5-26555. These observations are associated
    with program 10505.}
\altaffiltext{2}{Instituto de Astrof\'{i}sica de Canarias, La Laguna, Tenerife,
    Spain; ebernard@iac.es, monelli@iac.es, carme@iac.es, dio@iac.es,
    antapaj@iac.es, slhidalgo@iac.es.}
\altaffiltext{3}{Astronomical Institute, St. Petersburg State University,
    St. Petersburg, Russia.}
\altaffiltext{4}{Dominion Astrophysical Observatory, Herzberg Institute of
    Astrophysics, National Research Council, Victoria, Canada;
    peter.stetson@nrc-cnrc.gc.ca.}
\altaffiltext{5}{Departamento de Astrof\'{i}sica, Universidad de La Laguna,
    Tenerife, Spain.}
\altaffiltext{6}{INAF-Osservatorio Astronomico di Collurania,
    Teramo, Italy; cassisi@oa-teramo.inaf.it.}
\altaffiltext{7}{Department of Physics, Institut f\"ur Astronomie,
    ETH Z\"urich, Z\"urich, Switzerland; lucio@phys.ethz.ch.}
\altaffiltext{8}{Institut f\"ur Theoretische Physik, University of Zurich,
    Z\"urich, Switzerland; lucio@physik.unizh.ch.}
\altaffiltext{9}{School of Mathematics \& Physics, University of Tasmania,
    Hobart, Tasmania, Australia; andrew.cole@utas.edu.au.}
\altaffiltext{10}{Department of Astronomy, University of Minnesota,
    Minneapolis, USA; skillman@astro.umn.edu.}
\altaffiltext{11}{Kapteyn Astronomical Institute, University of Groningen,
    Groningen, Netherlands; etolstoy@astro.rug.nl.}

\begin{abstract}

 We present the first study of the variable star populations in the isolated
 dwarf spheroidal galaxies (dSph) Cetus and Tucana. Based on {\it Hubble Space
 Telescope} images obtained with the {\it Advanced Camera for Surveys} in the
 F475W and F814W bands, we identified 180 and 371 variables in Cetus and
 Tucana, respectively. The vast majority are RR~Lyrae stars. In Cetus we also
 found three anomalous Cepheids, four candidate binaries and one candidate
 long-period variable (LPV), while six anomalous Cepheids and seven LPV
 candidates were found in Tucana.
 Of the RR~Lyrae stars, 147 were identified as fundamental mode
 (RR$ab$) and only eight as first-overtone mode (RR$c$) in Cetus, with mean
 periods of 0.614 and 0.363 day, respectively. In Tucana we found 216
 RR$ab$ and 82 RR$c$ giving mean periods of 0.604 and 0.353 day.
 These values place both galaxies in the so-called Oosterhoff Gap, as is
 generally the case for dSph.
 We calculated the distance modulus to both galaxies using different
 approaches based on the properties of RR$ab$ and RR$c$, namely the
 luminosity-metallicity and period-luminosity-metallicity relations, and found
 values in excellent agreement with previous estimates using independent
 methods: (m$-$M)$_{0,Cet}$=24.46$\pm$0.12 and (m$-$M)$_{0,Tuc}$=24.74$\pm$0.12,
 corresponding to 780$\pm$40~kpc and 890$\pm$50~kpc.
 We also found numerous RR~Lyrae variables pulsating in both modes
 simultaneously (RR$d$): 17 in Cetus and 60 in Tucana.
 Tucana is, after Fornax, the second dSph in which such a large fraction of
 RR$d$ ($\sim$17\%) has been observed.
 We provide the photometry and pulsation parameters for all the variables, and
 compare the latter with values from the literature for well-studied dSph of
 the Local Group and Galactic globular clusters.

 The parallel WFPC2 fields were also searched for variables, as they lie well
 within the tidal radius of Cetus, and at its limit in the case of Tucana. No
 variables were found in the latter, while 15 were discovered in the outer
 field of Cetus (11 RR$ab$, 3 RR$c$ and 1 RR$d$), even though the lower
 signal-to-noise ratio of the observations did not allow us to measure their
 periods accurately. We provide their coordinates and approximate properties
 for completeness.

\end{abstract}

\keywords{galaxies: dwarf ---
          galaxies: individual (Cetus, Tucana) ---
          stars: horizontal-branch ---
          stars: variables: other ---
          Local Group}

\section{Introduction}

 Pulsating variable stars play a major role in the study of stellar
 populations and in cosmology, as their pulsational properties are
 traditionally used to determine distances and to put constraints on stellar
 physical properties. Because the pulsations occur at a particular phase of
 their evolution depending on the star mass, variable stars trace the spatial
 distribution of stellar populations of given ages, therefore highlighting the
 eventual radial trends across the studied galaxy \citep[e.g.,][]{gal04}.
 In addition, variations in the pulsational properties between individual
 stars of a particular type can trace subtle differences in the age and
 metallicity of the corresponding population
 \citep[e.g.,][hereafter Paper~I]{ber08}.

%%% Defines the alias to use 'Paper I' instead of 'Bernard et al. 2008'
    \defcitealias{ber08}{Paper~I}
    \defcitealias{har01}{Harbeck et al. 2001}

 To date, all adequate searches for RR Lyrae stars in dwarf galaxies have been
 successful, including some newly discovered ultra-faint Milky Way (MW)
 satellites \citep[e.g., Bo\"otes I, Canes Venatici I and
 II;][]{dal06,sie06,kue08,gre08}, satellites of M31 \citep[and references
 therein]{pri05} and isolated dwarfs \citep[e.g., IC\,1613,
 LGS\,3;][]{dol01,ber07}, independent of the galaxy morphological type.
 Their omnipresence, together with the wealth of information they can provide
 about their parent galaxies, makes the RR~Lyrae stars one of the best probe to
 the properties of the old populations.
 In particular, in the case of isolated galaxies which were not disturbed by
 interactions with massive galaxies, they represent a window to the processes
 of galaxy formation and to the internal mechanisms affecting their early
 evolution.

 With the goal of understanding these processes, we are carrying out a large
 project (LCID\footnotemark[12]) aiming at reconstructing the full star
 formation history of a sample of isolated dwarf galaxies of the Local Group
 (LG), based on very deep, multi-epoch {\it Hubble Space Telescope} ({\it HST})
 ACS data. The sample includes representatives of the three main morphological
 types---irregular, spheroidal and so-called transition dIrr/dSph---located
 further than about two virial radii from both the MW and M31. The project is
 described in more detail in a companion paper (Gallart et al. 2009, in
 preparation), and the first results concerning the star formation history
 (SFH) of Leo\,A were presented in \citet{col07}.

\footnotetext[12]{Local Cosmology from Isolated Dwarfs:
 http://www.iac.es/project/LCID/.}

 In the first paper of this series dedicated to the study of variable stars
 \citepalias{ber08}, we reported on the detection of old population gradients
 in the dSph galaxy Tucana from the properties of its RR~Lyrae stars.
 In the present paper, we focus our attention on the global population of
 variable stars in the two most isolated dwarf spheroidal galaxies known to
 date in the LG, namely Tucana and Cetus.
 An in-depth analysis of the properties of these variables and a comparison
 with the properties of the variables in the other galaxies of the LCID sample
 will be presented in a forthcoming paper.

\begin{figure}%[b]   % Fig. 1
% \epsscale{0.6} % preprint
\epsscale{1.25} % emulateapj
\plotone{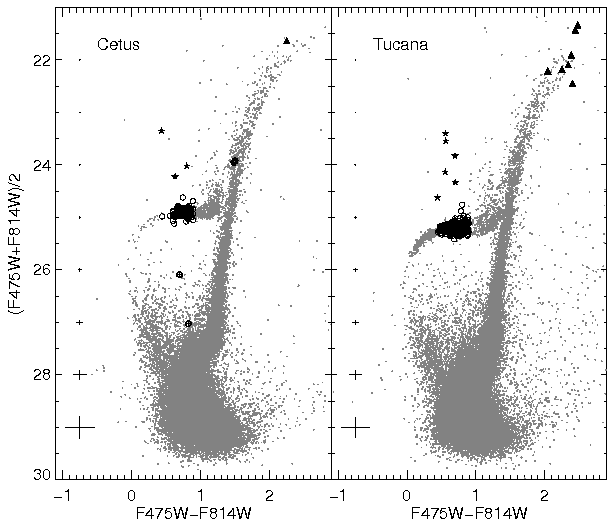}
\figcaption{Color-magnitude diagrams of Cetus and Tucana, where the candidate
 variables have been overplotted. Filled triangles, filled stars, open circles
 and crossed circles represent candidate LPV, above horizontal-branch
 variables, RR Lyrae stars and candidate binaries, respectively. The mean
 photometric error bars at given magnitudes are shown in each panel; note that
 they are smaller than the symbol size for magnitudes brighter than about 25.
\label{fig:1}}
\end{figure}

%%%%%%%%%%%%%%%%%%%%%%%%%%%%%%%%%%%%%%%%%%%%%%%%%%%%%%%%%%%%%%%%%%%%%%%%%%%%%%%%
%%%%%%%%%%%%%%%%%%%%%%%%%%%%%%%%%%%%%%%%%%%%%%%%%%%%%%%%%%%%%%%%%%%%%%%%%%%%%%%%
%%%
%%%   Tables -- Observing Logs
%%%

\begin{deluxetable}{ccccc}
\tabletypesize{\scriptsize}
\tablewidth{0pt}
\tablecaption{Observing Log for Cetus\label{tab1}}
\tablehead{
\colhead{Date} & \colhead{UT Start} & \colhead{MHJD\tablenotemark{a}} &
\colhead{Filter} & \colhead{Exp. Time}}
\startdata
 2006 Aug 28  &  11:32:20  &  53975.493956  &  F475W  &  1280 \\
 2006 Aug 28  &  11:56:35  &  53975.509789  &  F814W  &  1135 \\
 2006 Aug 28  &  13:06:18  &  53975.559326  &  F475W  &  1300 \\
 2006 Aug 28  &  13:30:53  &  53975.575449  &  F814W  &  1137 \\
 2006 Aug 28  &  14:44:46  &  53975.627590  &  F475W  &  1280 \\
 2006 Aug 28  &  15:09:01  &  53975.643585  &  F814W  &  1135 \\
 2006 Aug 28  &  16:18:10  &  53975.692566  &  F475W  &  1300 \\
 2006 Aug 28  &  16:42:45  &  53975.708689  &  F814W  &  1137 \\
 2006 Aug 28  &  17:56:22  &  53975.760645  &  F475W  &  1280 \\
 2006 Aug 28  &  18:20:37  &  53975.776480  &  F814W  &  1135 \\
 2006 Aug 28  &  19:31:50  &  53975.827057  &  F475W  &  1300 \\
 2006 Aug 28  &  19:56:25  &  53975.843180  &  F814W  &  1137 \\
 2006 Aug 29  &  01:55:57  &  53976.093689  &  F475W  &  1280 \\
 2006 Aug 29  &  02:20:13  &  53976.109696  &  F814W  &  1135 \\
 2006 Aug 29  &  03:30:37  &  53976.159545  &  F475W  &  1300 \\
 2006 Aug 29  &  03:55:13  &  53976.175679  &  F814W  &  1137 \\
 2006 Aug 29  &  05:07:47  &  53976.226906  &  F475W  &  1280 \\
 2006 Aug 29  &  05:32:02  &  53976.242901  &  F814W  &  1135 \\
 2006 Aug 29  &  06:42:26  &  53976.292751  &  F475W  &  1300 \\
 2006 Aug 29  &  07:07:01  &  53976.308874  &  F814W  &  1137 \\
 2006 Aug 29  &  08:19:38  &  53976.360135  &  F475W  &  1280 \\
 2006 Aug 29  &  08:43:53  &  53976.376130  &  F814W  &  1135 \\
 2006 Aug 29  &  09:54:16  &  53976.425968  &  F475W  &  1300 \\
 2006 Aug 29  &  10:18:51  &  53976.442091  &  F814W  &  1137 \\
 2006 Aug 29  &  11:31:29  &  53976.493364  &  F475W  &  1280 \\
 2006 Aug 29  &  11:55:44  &  53976.509359  &  F814W  &  1135 \\
 2006 Aug 29  &  13:06:05  &  53976.559174  &  F475W  &  1300 \\
 2006 Aug 29  &  13:30:40  &  53976.575141  &  F814W  &  1137 \\
 2006 Aug 29  &  16:19:15  &  53976.693202  &  F475W  &  1280 \\
 2006 Aug 29  &  16:43:30  &  53976.709197  &  F814W  &  1135 \\
 2006 Aug 29  &  17:53:48  &  53976.758977  &  F475W  &  1300 \\
 2006 Aug 29  &  18:18:23  &  53976.775100  &  F814W  &  1137 \\
 2006 Aug 30  &  05:07:08  &  53977.226453  &  F475W  &  1280 \\
 2006 Aug 30  &  05:31:23  &  53977.242449  &  F814W  &  1135 \\
 2006 Aug 30  &  06:41:04  &  53977.291800  &  F475W  &  1300 \\
 2006 Aug 30  &  07:05:39  &  53977.307772  &  F814W  &  1137 \\
 2006 Aug 30  &  08:18:58  &  53977.359671  &  F475W  &  1280 \\
 2006 Aug 30  &  08:43:12  &  53977.375654  &  F814W  &  1135 \\
 2006 Aug 30  &  09:52:52  &  53977.424995  &  F475W  &  1300 \\
 2006 Aug 30  &  10:17:26  &  53977.441106  &  F814W  &  1137 \\
 2006 Aug 30  &  11:31:22  &  53977.493281  &  F475W  &  1280 \\
 2006 Aug 30  &  11:55:37  &  53977.509127  &  F814W  &  1135 \\
 2006 Aug 30  &  13:04:41  &  53977.558200  &  F475W  &  1300 \\
 2006 Aug 30  &  13:29:16  &  53977.574323  &  F814W  &  1137 \\
 2006 Aug 30  &  14:43:03  &  53977.626395  &  F475W  &  1280 \\
 2006 Aug 30  &  15:07:18  &  53977.642390  &  F814W  &  1135 \\
 2006 Aug 30  &  16:16:30  &  53977.691406  &  F475W  &  1300 \\
 2006 Aug 30  &  16:41:05  &  53977.707529  &  F814W  &  1137 \\
 2006 Aug 30  &  17:52:27  &  53977.758154  &  F475W  &  1320 \\
 2006 Aug 30  &  18:17:22  &  53977.774276  &  F814W  &  1117
\enddata
\tablenotetext{a}{Modified Heliocentric Julian Date of mid-exposure: HJD$ - $2,400,000.}
\end{deluxetable}
    % Table 1

%%%
%%%   END Tables
%%%
%%%%%%%%%%%%%%%%%%%%%%%%%%%%%%%%%%%%%%%%%%%%%%%%%%%%%%%%%%%%%%%%%%%%%%%%%%%%%%%%
%%%%%%%%%%%%%%%%%%%%%%%%%%%%%%%%%%%%%%%%%%%%%%%%%%%%%%%%%%%%%%%%%%%%%%%%%%%%%%%%

\begin{figure*}%[b]   % Fig. 2
% \epsscale{1.0} % preprint
\epsscale{1.0} % emulateapj
\plottwo{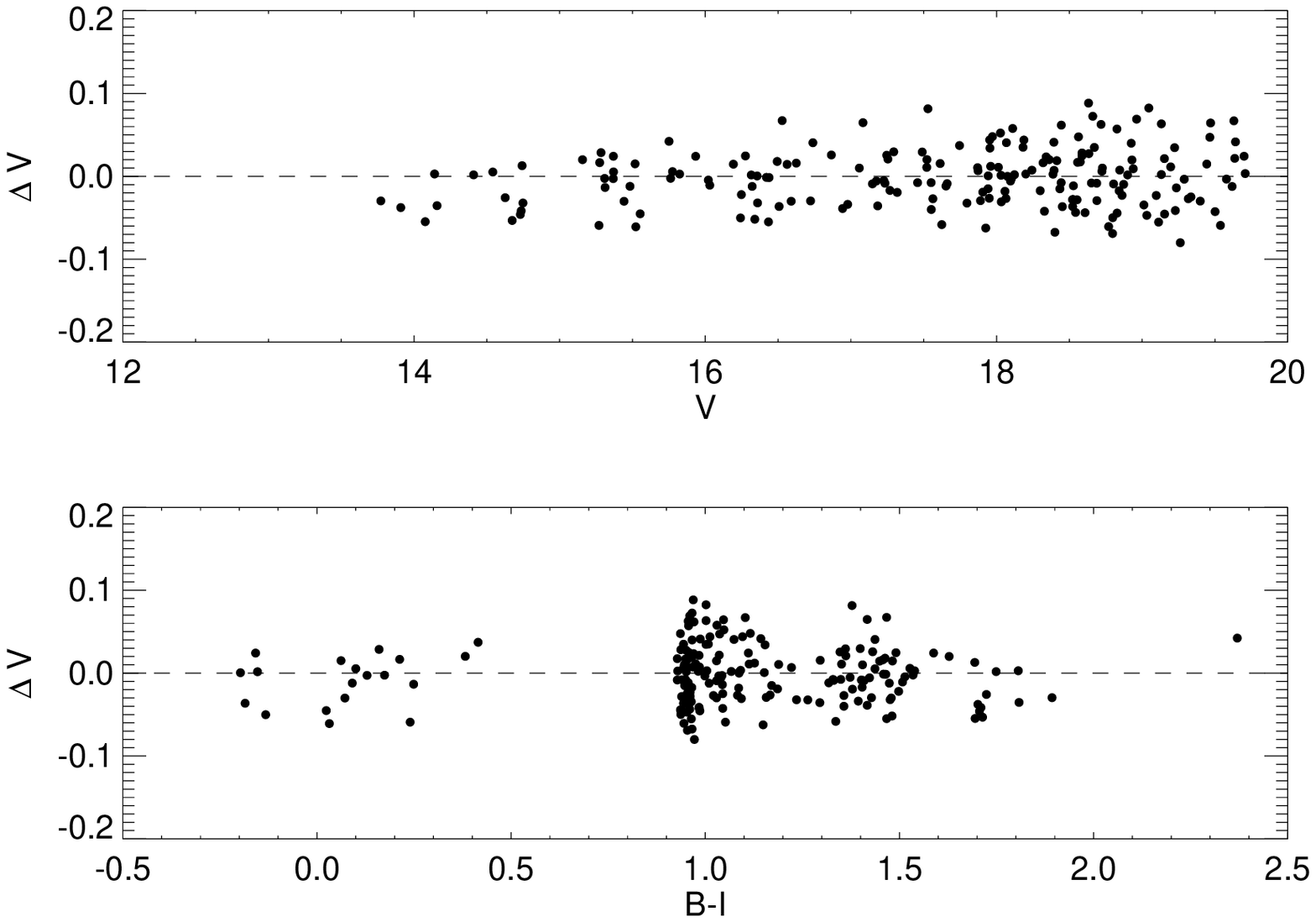}{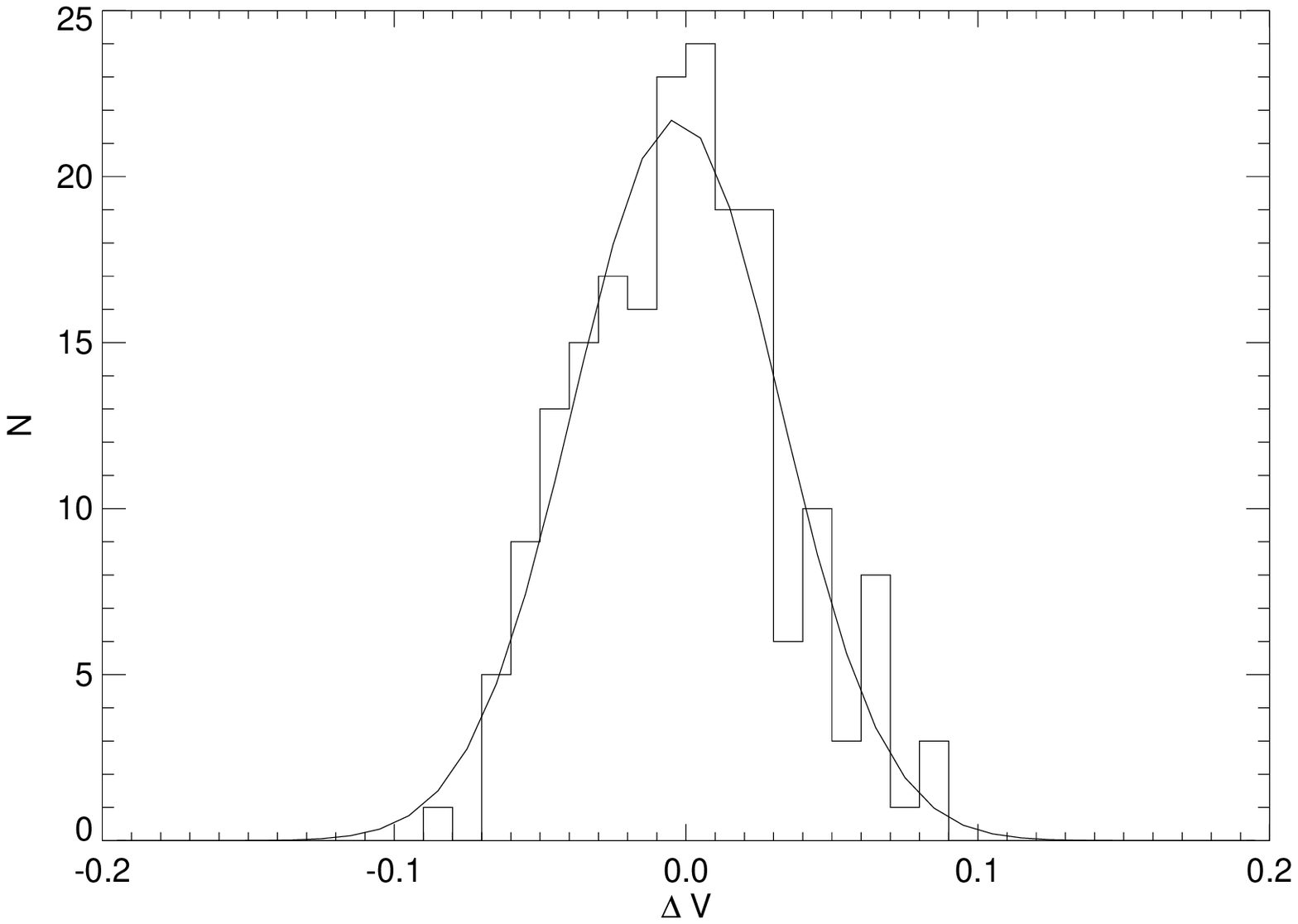}
\figcaption{{\it Left:} Residual distributions for the standardized {\it HST}
 sample of M92 stars as a function of magnitude and color.
 {\it Right:} The residual distribution is shown, overplotted with a Gaussian
 fit.
\label{fig:3}}
\end{figure*}

 Tucana was discovered as a likely LG dwarf spheroidal by \citet{lav90} based
 on resolved star CCD photometry. Later observations by \citet{lav92} confirmed
 its membership, who gave an upper limit on the distance modulus of 24.75, and
 yielded [Fe/H]=$-$1.9 from the mean metallicity-luminosity relation of the
 dwarf spheroidals of the LG. However, the color of the red giant branch (RGB)
 in deeper photometry from ground-based \citep{cas96} and {\it HST} WFPC2
 \citep{lav96} data indicate a metallicity as high as [Fe/H]=$-$1.6.
 In addition, \citet{cas96} estimated a metallicity spread of
 $\Delta$[Fe/H]$\sim$0.54 from the observed dispersion in color of the RGB,
 while \citet{har01} found a metallicity gradient as well as a bimodal [Fe/H]
 distribution using the horizontal branch (HB) morphology and the color of
 individual RGB stars, respectively.
 The hypothesis of multiple old populations in Tucana was strengthened in
 \citetalias{ber08} from the pulsational properties of its RR Lyrae stars.

 Cetus was discovered by \citet{whi99} upon visual inspection of southern sky
 survey plates. From follow-up observations, the authors derived a distance
 based on the tip of the RGB (TRGB) of $(m-M)_0$=24.45 $\pm$ 0.15, and
 [Fe/H]=$-$1.9 $\pm$ 0.2 from the color of the RGB. As in the case of Tucana,
 the color of the RGB from {\it HST} WFPC2 data gives a slightly more metal
 rich population with [Fe/H]=$-$1.7, and an intrinsic internal abundance
 dispersion of $\sim$0.2~dex \citep{sar02}.

 Because of their large distances from the MW, Cetus and Tucana have not
 previously been searched for variable stars, although the presence of RR Lyrae
 stars was suggested from the extension of the HB to the blue in both galaxies.
 \citet{cas96} reported the detection of three candidate long-period variables
 (LPV) near the TRGB of Tucana from their frame-to-frame photometric
 variations but no period search was attempted.

 In this paper we present the first in-depth analysis of the variable stars in
 Cetus and Tucana. A summary of the observations and data reduction is
 presented in section \ref{sec:2}, while \S~\ref{sec:3} and \S~\ref{sec:4} deal
 with the identification of variable stars and their completeness,
 respectively. In section \ref{sec:5} we describe the sample of RR~Lyrae stars
 at hand, and use their properties to estimate the distance of both galaxies in
 the following section. The remaining variables are presented in sections
 \ref{sec:7} and \ref{sec:8}.
 In the last section we discuss the results in the more general context of the
 properties of dwarf galaxies in the LG, and present our conclusions.

\section{Observations and Data Reduction}\label{sec:2}

\subsection{Primary ACS Imaging}

 The present analysis is based on observations obtained with the ACS onboard
 the {\it HST}. As the goal of these observations was to reach the oldest main
 sequence turn-offs with good signal-to-noise on the final photometry
 (S/N$>$10 at M$_I=+3$), we required 25 and 32 {\it HST} orbits for Cetus and
 Tucana, respectively. These were collected over about 2.5 and 5 consecutive
 days, between 2006 August 28 and 30 for Cetus and 2006 April 25 and 30 for
 Tucana. The observing sequence consisted of alternating $\sim$1100 seconds
 exposures in F475W and F814W for an optimal sampling of the light curves.
 The complete observing logs for Cetus and Tucana are given in
 Tables~\ref{tab1} and \ref{tab2}, respectively.

 The DAOPHOT/ALLFRAME suite of programs \citep{ste94} was used to obtain the
 instrumental photometry of the stars on the individual, non-drizzled images
 provided by the HST pipeline (the {\tt \_FLT} set). This pipeline carries out
 standard pre-reduction, including bias and dark subtraction, removal of the
 overscan regions, and flat fielding. Additionally, we used the pixel area maps
 and data quality masks to correct for the variations of the pixel areas on the
 sky around the field and to flag bad pixels.
 Standard calibration was carried out as described in \citet{sir05}, taking
 into account the updated zero-points of \citet{mac07} following the lowering
 of the Wide Field Channel temperature setpoint in July 2006. An in-depth
 description of the observations, data reduction and calibration, as well as
 results from extensive artificial star tests, are given in a companion paper
 (M. Monelli et al. 2009, in preparation). The final color-magnitude diagrams
 (CMDs) are shown in Fig.~\ref{fig:1}, where the $(F475W+F814W)/2 \sim V$
 filter combination was chosen for the ordinate axis so that the HB appears
 approximately horizontal.

\begin{figure}%[b]   % Fig. 3
% \epsscale{0.6} % preprint
\epsscale{1.2} % emulateapj
\plotone{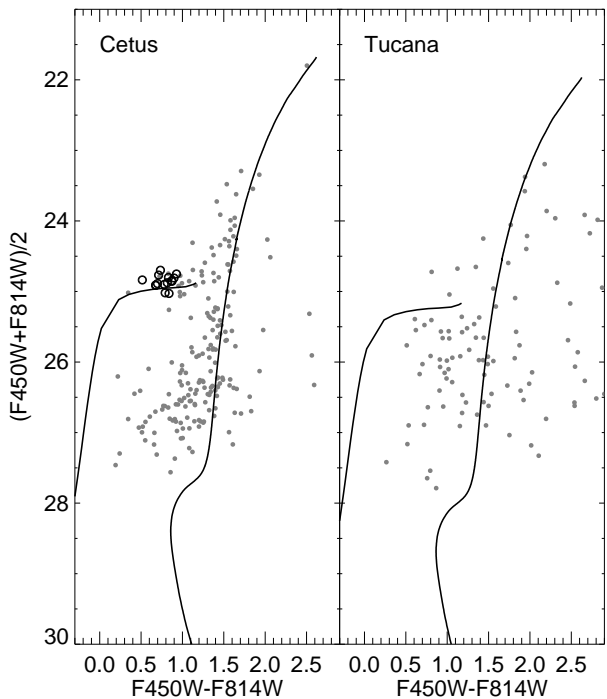}
\figcaption{Color-magnitude diagrams of the WFPC2 fields of Cetus and Tucana,
 where the RR Lyrae variables have been overplotted as open circles. The
 isochrones and zero-age horizontal branches are from the BaSTI library
 \citep[Z=0.0004, 13\,Gyr;][]{pie04}.
\label{fig:2}}
\end{figure}

%%%%%%%%%%%%%%%%%%%%%%%%%%%%%%%%%%%%%%%%%%%%%%%%%%%%%%%%%%%%%%%%%%%%%%%%%%%%%%%%
%%%%%%%%%%%%%%%%%%%%%%%%%%%%%%%%%%%%%%%%%%%%%%%%%%%%%%%%%%%%%%%%%%%%%%%%%%%%%%%%
%%%
%%%   Tables -- Observing Logs
%%%

\begin{deluxetable}{ccccc}
\tabletypesize{\scriptsize}
\tablewidth{0pt}
\tablecaption{Observing Log for Tucana\label{tab2}}
\tablehead{
\colhead{Date} & \colhead{UT Start} & \colhead{MHJD\tablenotemark{a}} &
\colhead{Filter} & \colhead{Exp. Time}}
\startdata
 2006 Apr 25  &  20:08:05  &  53850.848412  &  F475W  &  1070 \\
 2006 Apr 25  &  20:28:50  &  53850.862163  &  F814W  &  ~957 \\
 2006 Apr 25  &  21:42:13  &  53850.913900  &  F475W  &  1090 \\
 2006 Apr 25  &  22:03:18  &  53850.927894  &  F814W  &  ~979 \\
 2006 Apr 26  &  15:18:28  &  53851.647309  &  F475W  &  1070 \\
 2006 Apr 26  &  15:39:13  &  53851.661059  &  F814W  &  ~957 \\
 2006 Apr 26  &  16:52:34  &  53851.712773  &  F475W  &  1090 \\
 2006 Apr 26  &  17:13:39  &  53851.726767  &  F814W  &  ~979 \\
 2006 Apr 26  &  18:30:11  &  53851.780449  &  F475W  &  1070 \\
 2006 Apr 26  &  18:50:56  &  53851.794199  &  F814W  &  ~957 \\
 2006 Apr 26  &  20:04:17  &  53851.845913  &  F475W  &  1090 \\
 2006 Apr 26  &  20:25:22  &  53851.859907  &  F814W  &  ~979 \\
 2006 Apr 26  &  21:41:55  &  53851.913600  &  F475W  &  1070 \\
 2006 Apr 26  &  22:02:40  &  53851.927350  &  F814W  &  ~957 \\
 2006 Apr 26  &  23:16:00  &  53851.979053  &  F475W  &  1090 \\
 2006 Apr 26  &  23:37:05  &  53851.993046  &  F814W  &  ~979 \\
 2006 Apr 27  &  15:16:23  &  53852.645886  &  F475W  &  1070 \\
 2006 Apr 27  &  15:37:08  &  53852.661751  &  F814W  &  ~957 \\
 2006 Apr 27  &  16:50:29  &  53852.711350  &  F475W  &  1090 \\
 2006 Apr 27  &  17:11:34  &  53852.725344  &  F814W  &  ~979 \\
 2006 Apr 27  &  18:28:06  &  53852.779025  &  F475W  &  1070 \\
 2006 Apr 27  &  18:48:51  &  53852.792776  &  F814W  &  ~957 \\
 2006 Apr 27  &  20:02:11  &  53852.844478  &  F475W  &  1090 \\
 2006 Apr 27  &  20:23:16  &  53852.858472  &  F814W  &  ~979 \\
 2006 Apr 27  &  21:39:50  &  53852.912177  &  F475W  &  1070 \\
 2006 Apr 27  &  22:00:35  &  53852.925927  &  F814W  &  ~957 \\
 2006 Apr 27  &  23:13:55  &  53852.977630  &  F475W  &  1090 \\
 2006 Apr 27  &  23:35:00  &  53852.993727  &  F814W  &  ~979 \\
 2006 Apr 28  &  15:14:18  &  53853.644462  &  F475W  &  1070 \\
 2006 Apr 28  &  15:35:03  &  53853.660295  &  F814W  &  ~957 \\
 2006 Apr 28  &  16:48:20  &  53853.709880  &  F475W  &  1090 \\
 2006 Apr 28  &  17:09:25  &  53853.723874  &  F814W  &  ~979 \\
 2006 Apr 28  &  18:26:01  &  53853.777602  &  F475W  &  1070 \\
 2006 Apr 28  &  18:46:46  &  53853.791352  &  F814W  &  ~957 \\
 2006 Apr 28  &  20:00:03  &  53853.843020  &  F475W  &  1090 \\
 2006 Apr 28  &  20:21:08  &  53853.857013  &  F814W  &  ~979 \\
 2006 Apr 28  &  21:37:42  &  53853.910718  &  F475W  &  1070 \\
 2006 Apr 28  &  21:58:27  &  53853.924469  &  F814W  &  ~957 \\
 2006 Apr 28  &  23:11:45  &  53853.976148  &  F475W  &  1090 \\
 2006 Apr 28  &  23:32:50  &  53853.990141  &  F814W  &  ~979 \\
 2006 Apr 29  &  13:36:18  &  53854.576428  &  F475W  &  1070 \\
 2006 Apr 29  &  13:57:03  &  53854.592231  &  F814W  &  ~957 \\
 2006 Apr 29  &  15:10:18  &  53854.641823  &  F475W  &  1090 \\
 2006 Apr 29  &  15:31:23  &  53854.655816  &  F814W  &  ~979 \\
 2006 Apr 29  &  16:48:00  &  53854.709556  &  F475W  &  1070 \\
 2006 Apr 29  &  17:08:45  &  53854.723306  &  F814W  &  ~957 \\
 2006 Apr 29  &  18:22:00  &  53854.774951  &  F475W  &  1090 \\
 2006 Apr 29  &  18:43:05  &  53854.788944  &  F814W  &  ~979 \\
 2006 Apr 29  &  19:59:42  &  53854.842684  &  F475W  &  1070 \\
 2006 Apr 29  &  20:20:27  &  53854.856434  &  F814W  &  ~957 \\
 2006 Apr 29  &  21:33:43  &  53854.908090  &  F475W  &  1090 \\
 2006 Apr 29  &  21:54:48  &  53854.922084  &  F814W  &  ~979 \\
 2006 Apr 30  &  13:34:06  &  53855.574923  &  F475W  &  1070 \\
 2006 Apr 30  &  13:54:51  &  53855.588673  &  F814W  &  ~957 \\
 2006 Apr 30  &  15:08:04  &  53855.640294  &  F475W  &  1090 \\
 2006 Apr 30  &  15:29:09  &  53855.654288  &  F814W  &  ~979 \\
 2006 Apr 30  &  16:45:48  &  53855.708050  &  F475W  &  1070 \\
 2006 Apr 30  &  17:06:33  &  53855.721801  &  F814W  &  ~957 \\
 2006 Apr 30  &  18:19:45  &  53855.773411  &  F475W  &  1090 \\
 2006 Apr 30  &  18:40:50  &  53855.787404  &  F814W  &  ~979 \\
 2006 Apr 30  &  19:57:30  &  53855.841178  &  F475W  &  1070 \\
 2006 Apr 30  &  20:18:15  &  53855.854929  &  F814W  &  ~957 \\
 2006 Apr 30  &  21:31:27  &  53855.906539  &  F475W  &  1090 \\
 2006 Apr 30  &  21:52:32  &  53855.920532  &  F814W  &  ~979
\enddata
\tablenotetext{a}{Modified Heliocentric Julian Date of mid-exposure: HJD$ - $2,400,000.}
\end{deluxetable}
    % Table 2

%%%
%%%   END Tables
%%%
%%%%%%%%%%%%%%%%%%%%%%%%%%%%%%%%%%%%%%%%%%%%%%%%%%%%%%%%%%%%%%%%%%%%%%%%%%%%%%%%
%%%%%%%%%%%%%%%%%%%%%%%%%%%%%%%%%%%%%%%%%%%%%%%%%%%%%%%%%%%%%%%%%%%%%%%%%%%%%%%%

 The F475W and F814W magnitudes were transformed to Johnson $BVI$ to allow
 comparison with observations of variable stars in globular clusters and other
 galaxies reported in the literature. Given the large difference
 between these two photometric systems, the transformation was limited to the
 variable stars confined in the instability strip (IS), as their intermediate
 color and narrow temperature range minimized the difficulties related to
 extreme temperatures.
 To that purpose, we used our {\it HST} observations of NGC 6341 (M92), in
 which we found about 200 stars in common with Stetson's photometric standards
 in this cluster \citep{ste00}.
 Using linear regression, the transformation equations were determined to be:
\begin{eqnarray*}
B =& F475W + 0.07209(B-I) + 0.05126(B-I)^2 + 0.0075, \\
I =& F814W - 0.1030(B-I) + 0.06400(B-I)^2 + 0.0227, \\
V =& F475W - 0.7923(B-V) + 0.1983(B-V)^2 + 0.0070.
\end{eqnarray*}

 The transformations were then applied to our {\it HST} M92 data through an
 iterative process as described in \citet{sir05}, with the color terms being in
 the target photometric system.
 Figure~\ref{fig:3} shows the residuals between our {\it HST} `V' photometry
 and Stetson's standard V photometry, which have a median value of $-$0.001 and
 a standard deviation of 0.034. We obtain transformations of similar quality
 for the B and I bands, with median values of 0.004 and $-$0.002, and standard
 deviations of 0.032 and 0.030, respectively.
 The contribution of these transformations to the uncertainty of the individual
 magnitudes is therefore negligible ($\sim$0.002).

 Since pulsating variable stars undergo changes in effective temperature over
 the course of the pulsation cycle in addition to the changes in radius, their
 color varies accordingly. Thus, it is necessary to transform each phase point
 individually, taking into account the color of the star at that moment. This
 `instantaneous color' was obtained by using the consecutive F475W and F814W
 exposures observed within a single orbit.

\subsection{Parallel WFPC2 Imaging}

 These galaxies were also observed with the WFPC2 in the F450W and F814W bands
 as parallel exposures to the primary ACS observations, therefore providing the
 same number of observations with a similar exposure time in a second field in
 each galaxy. When preparing the observations, the orientation of the ACS field
 was chosen such that the parallel WFPC2 field would sample the outer regions
 of each galaxy of our sample.

 The images from the Wide Field chips were reduced individually as described in
 \citet{tur97}. On the other hand, we did not perform the photometry of the
 individual Planetary Camera (PC) images since they contain very few stars, and
 the large number of cosmic ray residuals prevented obtaining a reasonable
 registration. Instead, the photometry of the PC was performed on the averaged
 F450W and F814W images. The location of the stars from this chip on the CMDs
 indicates that no bright variables are expected (see below).

 The resulting photometry was calibrated to the flight system following the
 instructions from the HST Data Handbook for WFPC2.\footnotemark[13]

\footnotetext[13]{http://www.stsci.edu/instruments/wfpc2/Wfpc2\_dhb/wfpc2\_ ch52.html}

\begin{figure}%[t]   % Fig. 4
% \epsscale{0.7} % preprint
\epsscale{1.0} % emulateapj
\plotone{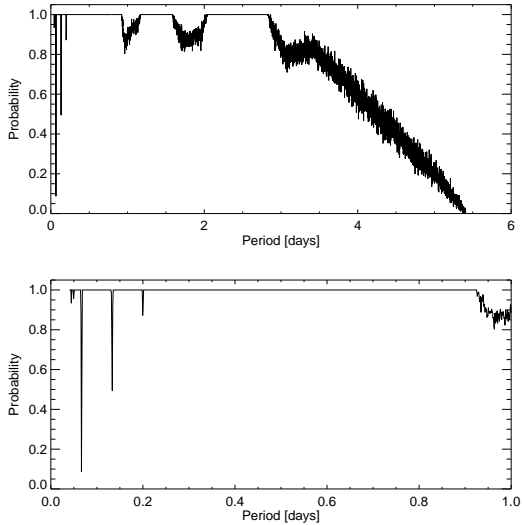}
\figcaption{Probability of detecting variable stars in Cetus as a function of
 period, for periods between about 1~hr and 6~days ({\it Top}), and close-up
 view for periods between about 1~hr and 1 day ({\it Bottom}).
\label{fig:4}}
\end{figure}

\begin{figure}%[t]   % Fig. 5
% \epsscale{0.7} % preprint
\epsscale{1.0} % emulateapj
\plotone{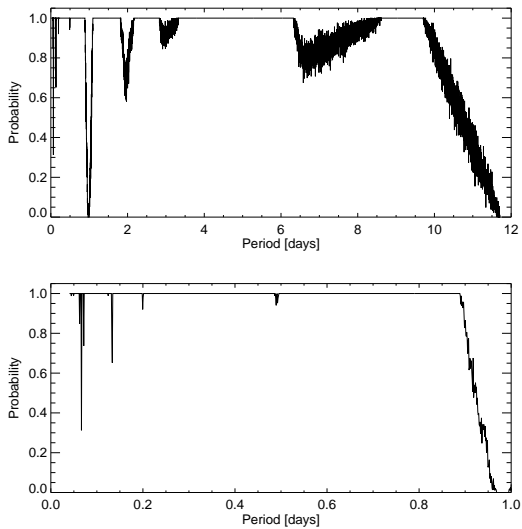}
\figcaption{Same as Fig.~\ref{fig:4} for Tucana.
\label{fig:5}}
\end{figure}

 Due to the lower sensitivity of the instrument with respect to the ACS,
 together with the smaller field of view and the peripheric location in the
 galaxies, very few stars were found in each field.
 The CMDs for both galaxies are shown in Fig.~\ref{fig:2}, where the isochrones
 and zero-age HB from the BaSTI library \citep{pie04} have been overplotted.
 For Cetus, we used $A_B$=0.123 and $A_I$=0.056 \citep{sch98} for the F450W and
 F814W bands, respectively, with an assumed dereddened distance modulus of
 24.46 (see Section~\ref{sec:6.3}). For Tucana, the BaSTI models were shifted
 assuming $A_B$=0.135, $A_I$=0.061, and (m$-$M)$_0$=24.74
 (Section~\ref{sec:6.3}).
 Figure~\ref{fig:2} shows that the CMD of the parallel WFPC2 field in Cetus is
 about 2.5 mag shallower than the ACS CMD, although it still presents a
 conspicuous RGB, and a red HB is visible.
 On the other hand, as expected from the much smaller tidal radius, less than a
 hundred objects were detected in Tucana, and the CMD does not present any
 recognizable features.

%%%%%%%%%%%%%%%%%%%%%%%%%%%%%%%%%%%%%%%%%%%%%%%%%%%%%%%%%%%%%%%%%%%%%%%%%%%%%%%%
%%%%%%%%%%%%%%%%%%%%%%%%%%%%%%%%%%%%%%%%%%%%%%%%%%%%%%%%%%%%%%%%%%%%%%%%%%%%%%%%
%%%
%%%   Tables -- Photometry
%%%

\begin{deluxetable}{ccc|ccc}
\tablewidth{0pt}
\tablecaption{Photometry of the Variable Stars in Cetus\label{tab3}}
\tablehead{
\colhead{MHJD\tablenotemark{a}} & \colhead{$m_{475}$} & \colhead{$\sigma_{475}$} &
\colhead{MHJD\tablenotemark{a}} & \colhead{$m_{814}$} & \colhead{$\sigma_{814}$}}
\startdata
\multicolumn{6}{c}{V001} \\ \hline
 53975.493956 & 25.065 & 0.031 &  53975.509789 & 24.339 & 0.035 \\
 53975.559326 & 25.206 & 0.036 &  53975.575449 & 24.400 & 0.048 \\
 53975.627590 & 25.388 & 0.039 &  53975.643585 & 24.433 & 0.124 \\
 53975.692566 & 25.454 & 0.044 &  53975.708689 & 24.299 & 0.099 \\
 53975.760645 & 25.510 & 0.034 &  53975.776480 & 24.543 & 0.044
\enddata
\tablecomments{Table \ref{tab3} is published in its entirety in the
electronic edition of the {\it Astrophysical Journal}.  A portion is
shown here for guidance regarding its form and content.}
\tablenotetext{a}{Modified Heliocentric Julian Date of mid-exposure: HJD$ - $2,400,000.}
\end{deluxetable}
    % Table 3

\begin{deluxetable}{ccc|ccc}
\tablewidth{0pt}
\tablecaption{Photometry of the Variable Stars in Tucana\label{tab4}}
\tablehead{
\colhead{MHJD\tablenotemark{a}} & \colhead{$m_{475}$} & \colhead{$\sigma_{475}$} &
\colhead{MHJD\tablenotemark{a}} & \colhead{$m_{814}$} & \colhead{$\sigma_{814}$}}
\startdata
\multicolumn{6}{c}{V001} \\ \hline
 53851.647309 & 25.990 & 0.065 &  53851.661059 & 25.232 & 0.053 \\
 53851.712773 & 26.005 & 0.037 &  53851.726767 & 25.186 & 0.055 \\
 53851.780449 & 25.420 & 0.045 &  53851.794199 & 24.935 & 0.047 \\
 53851.845913 & 25.157 & 0.045 &  53851.859907 & 24.823 & 0.059 \\
 53851.913600 & 25.508 & 0.053 &  53851.927350 & 24.859 & 0.052
\enddata
\tablecomments{Table \ref{tab4} is published in its entirety in the
electronic edition of the {\it Astrophysical Journal}.  A portion is
shown here for guidance regarding its form and content.}
\tablenotetext{a}{Modified Heliocentric Julian Date of mid-exposure: HJD$ - $2,400,000.}
\end{deluxetable}
    % Table 4

% \begin{deluxetable}{ccc|ccc|ccc}
\begin{deluxetable*}{ccc|ccc|ccc} % emulateapj

\tablewidth{0pt}
\tablecaption{Photometry of the Variable Stars in Cetus\label{tab5}}
\tablehead{
\colhead{MHJD\tablenotemark{a}} & \colhead{$B$} & \colhead{$\sigma_B$} &
\colhead{MHJD\tablenotemark{a}} & \colhead{$V$} & \colhead{$\sigma_V$} &
\colhead{MHJD\tablenotemark{a}} & \colhead{$I$} & \colhead{$\sigma_I$}}
\startdata
\multicolumn{9}{c}{V001} \\ \hline
 53975.493956 & 25.171 & 0.031 &  53975.501872 & 24.816 & 0.031 &  53975.509789 & 24.320 & 0.035 \\
 53975.559326 & 25.327 & 0.036 &  53975.567387 & 24.929 & 0.036 &  53975.575449 & 24.382 & 0.048 \\
 53975.627590 & 25.541 & 0.039 &  53975.635588 & 25.061 & 0.039 &  53975.643585 & 24.421 & 0.124 \\
 53975.692566 & 25.653 & 0.044 &  53975.700628 & 25.062 & 0.044 &  53975.708689 & 24.299 & 0.099 \\
 53975.760645 & 25.665 & 0.034 &  53975.768563 & 25.179 & 0.034 &  53975.776480 & 24.531 & 0.044
\enddata
\tablecomments{Table \ref{tab5} is published in its entirety in the
electronic edition of the {\it Astrophysical Journal}.  A portion is
shown here for guidance regarding its form and content.}
\tablenotetext{a}{Modified Heliocentric Julian Date of mid-exposure: HJD$ - $2,400,000.}
    % Table 5
% \end{deluxetable}
\end{deluxetable*}               % emulateapj

% \begin{deluxetable}{ccc|ccc|ccc}
\begin{deluxetable*}{ccc|ccc|ccc} % emulateapj

\tablewidth{0pt}
\tablecaption{Photometry of the Variable Stars in Tucana\label{tab6}}
\tablehead{
\colhead{MHJD\tablenotemark{a}} & \colhead{$B$} & \colhead{$\sigma_B$} &
\colhead{MHJD\tablenotemark{a}} & \colhead{$V$} & \colhead{$\sigma_V$} &
\colhead{MHJD\tablenotemark{a}} & \colhead{$I$} & \colhead{$\sigma_I$}}
\startdata
\multicolumn{9}{c}{V001} \\ \hline
 53851.647309 & 26.102 & 0.065 &  53851.654184 & 25.730 & 0.065 &  53851.661059 & 25.214 & 0.053 \\
 53851.712773 & 26.129 & 0.037 &  53851.719770 & 25.723 & 0.037 &  53851.726767 & 25.169 & 0.055 \\
 53851.780449 & 25.485 & 0.045 &  53851.787324 & 25.257 & 0.045 &  53851.794199 & 24.920 & 0.047 \\
 53851.845913 & 25.200 & 0.045 &  53851.852910 & 25.049 & 0.045 &  53851.859907 & 24.816 & 0.059 \\
 53851.913600 & 25.600 & 0.053 &  53851.920475 & 25.286 & 0.053 &  53851.927350 & 24.840 & 0.052
\enddata
\tablecomments{Table \ref{tab6} is published in its entirety in the
electronic edition of the {\it Astrophysical Journal}.  A portion is
shown here for guidance regarding its form and content.}
\tablenotetext{a}{Modified Heliocentric Julian Date of mid-exposure: HJD$ - $2,400,000.}
    % Table 6
% \end{deluxetable}
\end{deluxetable*}               % emulateapj

%%%
%%%   END Tables
%%%
%%%%%%%%%%%%%%%%%%%%%%%%%%%%%%%%%%%%%%%%%%%%%%%%%%%%%%%%%%%%%%%%%%%%%%%%%%%%%%%%
%%%%%%%%%%%%%%%%%%%%%%%%%%%%%%%%%%%%%%%%%%%%%%%%%%%%%%%%%%%%%%%%%%%%%%%%%%%%%%%%

\section{Identification of Variable Stars}\label{sec:3}

 The candidate variables were extracted from the ACS photometry using the
 Welch-Stetson variability index \citep{wel93}, which makes optimal use of our
 alternating F475W and F814W measurements by employing the correlation in
 brightness change in paired frames. This process yielded 379 and 565
 candidates in the primary field of Cetus and Tucana, respectively.
 A preliminary check of the light-curve and position on the CMD, together with
 a careful inspection of the stacked image, allowed us to discard false
 detections due to cosmic-ray hits, chip defects or stars located under the
 wings of bright stars.

\begin{figure*}%[t]   % Fig. 6
% \epsscale{1.0} % preprint
\epsscale{0.9} % emulateapj
\plotone{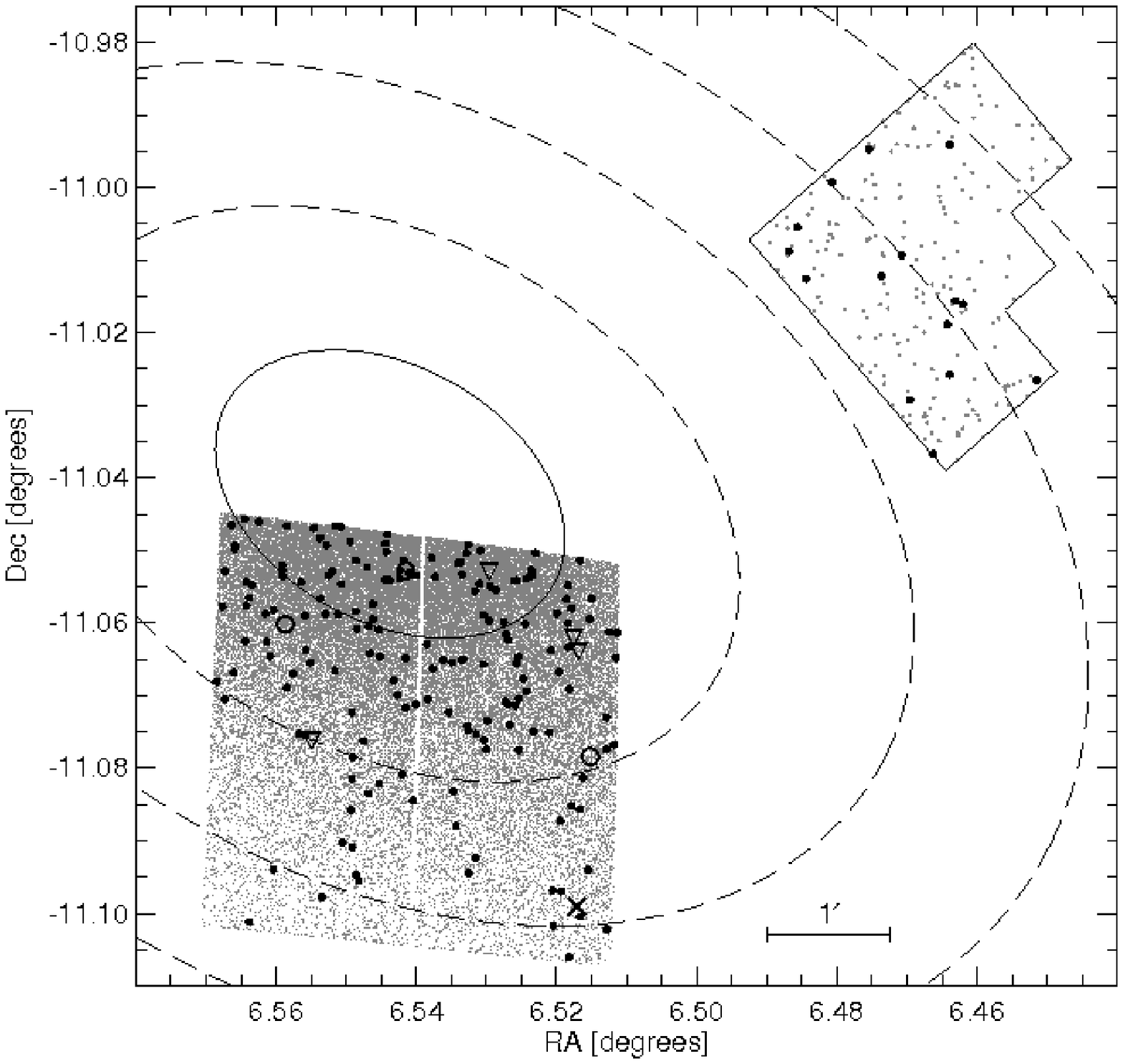}
\figcaption{Spatial distribution of stars in the ACS and WFPC2 fields in Cetus.
 RR Lyrae stars ({\it larger dots}), Cepheids ({\it open circles}), and
 candidate binaries ({\it open triangles}) and LPV ({\it cross}) are also
 shown. Solid- and dashed-line ellipses represent the core radius
 ($r_c=1.3\arcmin \pm 0.1$) and integer multiples of the core radius, from
 \citet{mcc06}.
\label{fig:6}}
\end{figure*}

 The period search was first performed on the suspected variables through both
 Fourier analysis following the prescription of \citet{hor86}, and the
 phase-dispersion minimization algorithm \citep{ste78}, both methods taking
 into account the information from both bands simultaneously.
 As both methods were giving very similar periods, the former was chosen over
 the latter for its swiftness. For each variable, datapoints with error bars
 larger than 3-$\sigma$ above the mean error bar size were rejected through
 sigma clipping with five iterations. As the period-finding program is
 interactive, it was possible to selectively reject more or less datapoints
 depending on the light curve quality before recalculating the periodogram.
 Except in a few particular cases (e.g., when a variable was located close to a
 bad column or on the border of a chip, thus yielding several very discrepant
 datapoints, i.e., V39 and V135 in Cetus; V51, V80, and V137 in Tucana), we
 found that the period-search was not affected by a few bad points.
 The periods were then refined by hand upon visual inspection of the light
 curves in both bands simultaneously. Given the short timebase of the
 observations ($\la$6~days), the periods are given with three significant
 figures only.
 The accuracy of period evaluations was estimated from the degradation of the
 quality of the light-curves when applying small offsets. It mainly depends on
 the period itself and on the time interval covered by observations, and ranges
 from about 0.001 day for the shorter period RR Lyrae stars to few hundredths
 of a day for the longest period Cepheids in Cetus.
 We ended up with 180 variables in Cetus and 371 in Tucana. These
 are shown in Fig.~\ref{fig:1} overplotted on the CMD of each galaxy using
 their intensity-averaged magnitudes (see below). The individual F475W and
 F814W measurements for all of the variables are listed in Tables~\ref{tab3}
 (Cetus) and \ref{tab4} (Tucana), while Tables~\ref{tab5} and \ref{tab6} give
 the transformed B, V, and I magnitudes as described above. The Julian date for
 the V band is simply the midpoint between consecutive F475W and F814W
 observations.

 The classification of the candidates was based on their light-curve morphology
 and position in the CMD.
 In Cetus, we found 172 RR Lyrae stars, three above horizontal-branch variables
 (AHB), four candidate binaries and a candidate LPV, while in Tucana we found
 358 RR Lyrae, six AHB and seven LPV candidates around the TRGB. Because of the
 short timebase and relatively small number of observations, no attempt was
 made to find a period for the binaries and LPV.

 To obtain the amplitudes and intensity-averaged magnitudes of the monoperiodic
 variables in the IS, we fitted the light-curves with a set of templates partly
 based on the set of \citet{lay99}. Two other templates---for Cepheids, one
 sawtooth-like and one with constant-light at minimum---were built by averaging
 the light-curves of ten well-measured IC\,1613 Cepheids from the OGLE database
 \citep{uda01}.
 On the other hand, the amplitude of the double-mode RR Lyrae stars was
 measured from a low-order Fourier fit to the light-curve phased with the
 primary period after prewhitening of the secondary period.
 The mean magnitude and color of the RR$d$ and candidate binaries and LPV are
 weighted averages, and are therefore only approximate. Tables
 \ref{tab7} and \ref{tab8} summarize the properties of the variable stars in
 the ACS field of each galaxy. The first and second columns give the
 identification number and variable type, while the next two list the
 equatorial coordinates (J2000.0). Columns (5) and (6) give the primary period
 in days, i.e., the first-overtone period in the case of the RR$d$, and the
 logarithm of this period. The intensity-averaged magnitudes
 $\langle F475W \rangle$ and $\langle F814W \rangle$, and color
 $\langle F475W \rangle- \langle F814W \rangle$ are given in columns (7), (9),
 and (11), and the amplitudes in the F475W and F814W bands measured from the
 template fits are listed in the eighth and tenth columns. The last six columns
 alternately list the intensity-averaged magnitudes and amplitudes in the
 Johnson B, V, and I bands. Approximate values are listed in italics.

 The same procedure was followed with the WFPC2 photometry of both galaxies.
 As expected from the appearance of the WFPC2 CMD, no candidate variables were
 detected in Tucana. In the outer field of Cetus, we found 15 RR~Lyrae
 variables. However, the low signal-to-noise at the magnitude of the HB
 produced rather noisy light-curves, and some low amplitude variables might
 have been missed.
 Given the small number of variables in the parallel field and the generally
 lower quality of their photometry and inferred parameters, we only give
 their coordinates and approximate parameters in Table~\ref{tab9} for
 completeness. These variables will not be taken into account when calculating
 average properties.

\section{Completeness}\label{sec:4}

\begin{figure}%[b]   % Fig. 7
% \epsscale{0.7} % preprint
\epsscale{1.2} % emulateapj
\plotone{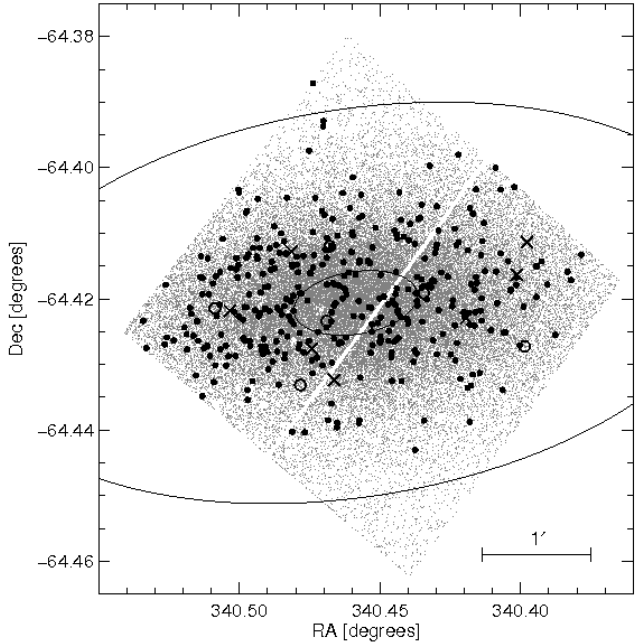}
\figcaption{Same as Fig.~\ref{fig:6}, but for Tucana.
 The ellipses represent the core and tidal radii at $r_c=0.59\arcmin \pm 0.01$
 and $r_t=3.87\arcmin \pm 0.37$.
\label{fig:7} }
\end{figure}

 Given that the properties of the variable stars (e.g., $\langle P_{ab}
 \rangle$, f$_c=N_c / (N_{ab}+N_c)$, specific frequency of anomalous Cepheids)
 are often used to compare stellar systems, and that these properties can vary
 within a single one of these systems \citepalias[see, e.g.,][]{ber08}, it is
 necessary to estimate the completeness of the sample of variable stars at
 hand. Incompleteness can be due to several reasons, mainly stellar crowding
 and signal-to-noise (SN) limitations, temporal sampling, and spatial coverage.

 In this particular case, the high spatial resolution of the ACS and the depth
 of our data imply that incompleteness will become noticeable only well below
 the HB. Artificial-star tests (see M. Monelli et al. 2009, in preparation)
 indicate that the completeness is higher than 97\% at (F475W+F814W)/2 $\sim$
 25.5 and 26.0 for Cetus and Tucana, respectively. Therefore, down to these
 magnitudes only variables with amplitudes smaller than the error bars at this
 magnitude ($\sim$0.1) might have been missed.
 Crowding and low SN only limit our ability to detect variables fainter than
 the HB, e.g., SX Phoenicis stars and binary systems. In addition, even though
 these variables are most likely present in both galaxies, the relatively long
 exposure time smoothing out the variations in luminosity and the rather slow
 temporal sampling precluded the detection of these short-period variables.

 It is also important to estimate the completeness due to temporal sampling,
 since the actual time distribution of the observations can affect the
 detection of variables with given periods more than others.
 To this purpose, we carried out numerical simulations similar to those
 described in \citet{ber02}.
 Basically, we simulate a large number of variable stars (one million here)
 with periods randomly distributed between about one hour (0.04 day) and
 12~days, and random phases.
 Given that the period of the real stars was searched on the data from both
 bands simultaneously, for these tests we also combined together the
 observation times of both bands, which were folded according to the period
 and initial phase of each artificial variable star.
 We then counted the number of observations at given phase ranges. A variable
 star was considered recovered if it fulfills the following criteria:
 {\it (a)} at least two observations around maximum light ($\phi=0. \pm 0.1$);
 {\it (b)} at least two phase points during descending light ($0.2<\phi<0.5$);
 and {\it (c)} at least three phase points during minimum light
 ($0.5<\phi<0.8$).

 We then calculated the detection probability as a function of period as the
 fraction of artificial stars in period intervals of 0.001~day fulfilling the
 criteria given above. The result is presented in Fig.~\ref{fig:4} and
 \ref{fig:5} for Cetus and Tucana, respectively.
 In the upper panel of each figure, the maximum period that is displayed is
 limited by the observational timebase of each galaxy, given the above
 criteria. Note, however, that while it is common to detect variables with
 periods longer than the time-span of the observations, it is not possible to
 determine their periods.
 For periods shorter than about 3.5 days, one can see that the
 spectrum of each galaxy displays similar features. These are due to the
 observational strategy, constrained by the orbital period of the {\it HST}.
 For example, the minimum on the lower panel of each figure at period
 $\sim$0.065 day corresponds to the shortest time between two consecutive
 images of the same band.

 The main difference appears on the spectrum of Tucana at P$\sim$1~day, and is
 the consequence of the location of Tucana in the sky, together with the
 presence of the South-Atlantic Anomaly (SAA). The {\tt SAME ORIENT}
 requirement did not allow these observations to be scheduled at times when the
 passages of {\it HST} through the SAA could be hidden within the target's
 Earth occultations. This, in turn, forced to clump the observations together
 at roughly the same time of the day each day. Cetus, on the other hand, was
 not affected by the SAA and has no `blind' period.

 Note that these detection probabilities should be taken as a lower limit,
 since several variables were observed that do not fulfill one (or more) of the
 above criteria (e.g., V134 in Cetus, V125 in Tucana), or variables that have
 periods longer than 12~days (candidate LPV in both galaxies).
 Moreover, variables with periods of about one day would most likely be
 Cepheids. In Tucana, there are very few stars in the IS above the HB, and any
 variable with amplitude greater than the error bars at a given magnitude would
 be detected. Removing the second criteria (in itself not necessary to
 {\it detect} a variable) also effectively removes the blind period of Tucana
 at P$\sim$1~day; the minimum probability at P=0.97~day rises to $\sim$0.2.
 In any case, for both galaxies the probability to detect a variable star in
 the period range of RR~Lyrae stars ($\sim$ 0.2--0.8 day) is basically one.

\begin{figure}%[b]   % Fig. 8
% \epsscale{0.6} % preprint
\epsscale{0.80} % emulateapj
\plotone{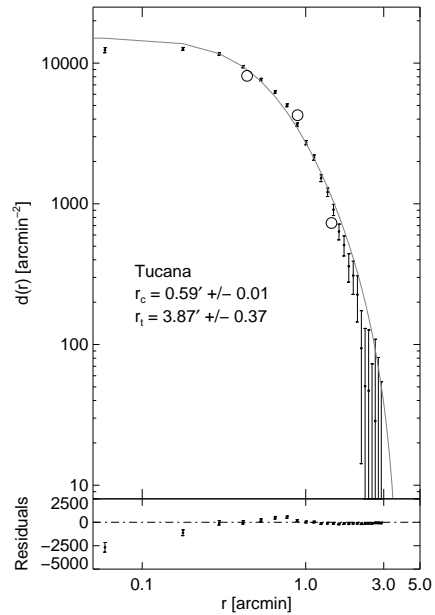}
\figcaption{Radial stellar profile of Tucana ({\it dots}) and the fitted King
 profile ({\it solid line}). The error bars take into account the Poisson
 uncertainty in the star counts and the uncertainty in the background estimate.
 The larger circles show the stellar profile of the RR~Lyrae stars, shifted
 vertically to fit on the profile of the non-variable stars.
\label{fig:8}}
\end{figure}

%%%%%%%%%%%%%%%%%%%%%%%%%%%%%%%%%%%%%%%%%%%%%%%%%%%%%%%%%%%%%%%%%%%%%%%%%%%%%%%%
%%%%%%%%%%%%%%%%%%%%%%%%%%%%%%%%%%%%%%%%%%%%%%%%%%%%%%%%%%%%%%%%%%%%%%%%%%%%%%%%
%%%
%%%   Tables -- Properties of variable stars
%%%

% \begin{deluxetable}{cccccccccccccccccc}
\begin{deluxetable*}{cccccccccccccccccc} % emulateapj
\tabletypesize{\scriptsize}
% \rotate

\tablewidth{0pt}
\tablecaption{Pulsation Properties of Variable Stars in Cetus -- ACS field\tablenotemark{a}\label{tab7}}
\tablehead{
\colhead{} & \colhead{} & \colhead{R.A.} & \colhead{Decl.} & \colhead{Period} &
\colhead{} & \colhead{} & \colhead{} & \colhead{} & \colhead{} &
\colhead{$\langle F475W \rangle -$}\\
\colhead{ID} & \colhead{Type} & \colhead{(J2000)} & \colhead{(J2000)} &
\colhead{(days)} & \colhead{log P} &
\colhead{$\langle F475W \rangle$} & \colhead{$A_{475}$} &
\colhead{$\langle F814W \rangle$} & \colhead{$A_{814}$} &
\colhead{$\langle F814W \rangle$} &
\colhead{$\langle B \rangle$} & \colhead{$A_B$} &
\colhead{$\langle V \rangle$} & \colhead{$A_V$} &
\colhead{$\langle I \rangle$} & \colhead{$A_I$}}
\startdata
 V001 & $ab$ &  00 26 02.73 & $-$11 03 40.6 &      0.664  &      $-$0.178  &      25.308  &      0.624   &      24.465  &      0.354   &      0.844  &      25.438  &      0.683  &      25.023  &      0.562  &      24.450  &      0.378  \\
 V002 &  $d$ &  00 26 02.77 & $-$11 03 53.2 & {\it 0.386} & {\it $-$0.413} & {\it 25.281} & {\it 0.828}  & {\it 24.571} & {\it 0.509}  & {\it 0.710} & {\it 25.379} & {\it 0.911} & {\it 25.045} & {\it 0.671} & {\it 24.555} & {\it 0.507} \\
 V003 & $ab$ &  00 26 02.84 & $-$11 04 36.3 &      0.591  &      $-$0.228  &      25.317  &      1.150   &      24.533  &      0.600   &      0.783  &      25.431  &      1.264  &      25.059  &      0.841  &      24.516  &      0.610  \\
 V004 & $ab$ &  00 26 03.00 & $-$11 03 40.4 &      0.638  &      $-$0.195  &      25.343  &      0.569   &      24.493  &      0.274   &      0.849  &      25.473  &      0.641  &      25.055  &      0.475  &      24.478  &      0.281  \\
 V005 & $ab$ &  00 26 03.07 & $-$11 06 08.2 &      0.601  &      $-$0.221  &      25.343  &      1.069   &      24.568  &      0.468   &      0.775  &      25.459  &      1.186  &      25.083  &      0.857  &      24.556  &      0.466
\enddata
\tablenotetext{a}{Approximate values are listed in italics.}
\tablecomments{Table \ref{tab7} is published in its entirety in the
electronic edition of the {\it Astrophysical Journal}.  A portion is
shown here for guidance regarding its form and content.}
    % Table 7
% \end{deluxetable}
\end{deluxetable*}               % emulateapj

% \begin{deluxetable}{cccccccccccccccccc}
\begin{deluxetable*}{cccccccccccccccccc} % emulateapj
\tabletypesize{\scriptsize}
% \rotate

\tablewidth{0pt}
\tablecaption{Pulsation Properties of Variable Stars in Tucana\tablenotemark{a}\label{tab8}}
\tablehead{
\colhead{} & \colhead{} & \colhead{R.A.} & \colhead{Decl.} & \colhead{Period} &
\colhead{} & \colhead{} & \colhead{} & \colhead{} & \colhead{} &
\colhead{$\langle F475W \rangle -$}\\
\colhead{ID} &  \colhead{Type} &  \colhead{(J2000)} & \colhead{(J2000)} &
\colhead{(days)} & \colhead{log P} &
\colhead{$\langle F475W \rangle$} & \colhead{$A_{475}$} &
\colhead{$\langle F814W \rangle$} & \colhead{A$_{814}$} &
\colhead{$\langle F814W \rangle$} &
\colhead{$\langle B \rangle$} & \colhead{$A_B$} &
\colhead{$\langle V \rangle$} & \colhead{$A_V$} &
\colhead{$\langle I \rangle$} & \colhead{$A_I$}}
\startdata
 V001 &  $d$ &  22 41 30.84 & $-$64 24 47.8 & {\it 0.370} & {\it $-$0.432} & {\it 25.651} & {\it 0.958}  & {\it 24.972} & {\it 0.523}  & {\it 0.679} & {\it 25.749} & {\it 1.058} & {\it 25.429} & {\it 0.757} & {\it 24.955} & {\it 0.513} \\
 V002 &  $c$ &  22 41 31.70 & $-$64 25 01.7 &      0.341  &      $-$0.467  &      25.599  &      0.650   &      24.992  &      0.321   &      0.607  &      25.689  &      0.717  &      25.389  &      0.523  &      24.976  &      0.318  \\
 V003 & $ab$ &  22 41 32.00 & $-$64 24 57.2 &      0.627  &      $-$0.203  &      25.495  &      1.089   &      24.769  &      0.469   &      0.727  &      25.620  &      1.132  &      25.222  &      1.001  &      24.751  &      0.472  \\
 V004 &  $c$ &  22 41 33.05 & $-$64 25 31.9 &      0.372  &      $-$0.429  &      25.556  &      0.528   &      24.888  &      0.308   &      0.667  &      25.653  &      0.559  &      25.325  &      0.462  &      24.870  &      0.325  \\
 V005 & $ab$ &  22 41 33.74 & $-$64 25 04.2 &      0.543  &      $-$0.265  &      25.613  &      1.359   &      24.939  &      0.684   &      0.674  &      25.725  &      1.464  &      25.364  &      1.152  &      24.923  &      0.684
\enddata
\tablenotetext{a}{Approximate values are listed in italics.}
\tablecomments{Table \ref{tab8} is published in its entirety in the
electronic edition of the {\it Astrophysical Journal}.  A portion is
shown here for guidance regarding its form and content.}
    % Table 8
% \end{deluxetable}
\end{deluxetable*}               % emulateapj

% \begin{deluxetable}{cccccccccccc}
\begin{deluxetable*}{cccccccccccc} % emulateapj
\tabletypesize{\scriptsize}

\tablewidth{0pt}
\tablecaption{Pulsation Properties of Variable Stars in Cetus -- WFPC2 field\tablenotemark{a}\label{tab9}}
\tablehead{
\colhead{ } & \colhead{ } & \colhead{R.A.} & \colhead{Decl.} & \colhead{Period} &
\colhead{} & \colhead{} & \colhead{} & \colhead{$\langle F450W \rangle -$}\\
\colhead{ID} & \colhead{Type} & \colhead{(J2000)} & \colhead{(J2000)} &
\colhead{(days)} & \colhead{log P} &
\colhead{$\langle F450W \rangle$} & \colhead{$\langle F814W \rangle$} &
\colhead{$\langle F814W \rangle$} &
\colhead{$A_{450}$} & \colhead{$A_{814}$}}
\startdata
 V181 & $ab$ &  00 25 48.36 & $-$11 01 35.6 &      0.74   &      $-$0.13   &      25.216  &      24.288  &      0.928  &      0.950  &      0.329   \\
 V182 & $ab$ &  00 25 50.89 & $-$11 00 57.8 &      0.58   &      $-$0.24   &      25.415  &      24.624  &      0.792  &      0.866  &      0.497   \\
 V183 &  $c$ &  00 25 51.15 & $-$11 00 56.2 &      0.41   &      $-$0.39   &      25.238  &      24.539  &      0.700  &      0.625  &      0.317   \\
 V184 & $ab$ &  00 25 51.34 & $-$11 01 32.7 &      0.59   &      $-$0.23   &      25.447  &      24.609  &      0.838  &      0.921  &      0.447   \\
 V185 & $ab$ &  00 25 51.34 & $-$10 59 38.7 &      0.62   &      $-$0.21   &      25.124  &      24.412  &      0.713  &      0.447  &      0.351   \\
 V186 &  $c$ &  00 25 51.42 & $-$11 01 07.9 &      0.39   &      $-$0.41   &      25.252  &      24.576  &      0.676  &      0.634  &      0.202   \\
 V187 &  $d$ &  00 25 51.90 & $-$11 02 12.2 & {\it 0.4}   & {\it $-$0.4}   & {\it 25.235} & {\it 24.477} & {\it 0.758} & {\it 0.467} & {\it 0.295}  \\
 V188 & $ab$ &  00 25 52.69 & $-$11 01 45.2 &      0.71   &      $-$0.15   &      25.283  &      24.413  &      0.870  &      0.729  &      0.404   \\
 V189 &  $c$ &  00 25 52.99 & $-$11 00 33.7 &      0.353  &      $-$0.452  &      25.094  &      24.579  &      0.515  &      0.745  &      0.318   \\
 V180 & $ab$ &  00 25 53.67 & $-$11 00 43.9 &      0.56   &      $-$0.25   &      25.247  &      24.570  &      0.677  &      1.130  &      0.635   \\
 V191 & $ab$ &  00 25 54.10 & $-$10 59 41.1 &      0.62   &      $-$0.21   &      25.285  &      24.468  &      0.818  &      0.693  &      0.486   \\
 V192 & $ab$ &  00 25 55.36 & $-$10 59 57.3 &      0.59   &      $-$0.23   &      25.066  &      24.332  &      0.734  &      1.531  &      0.467   \\
 V193 & $ab$ &  00 25 56.26 & $-$11 00 45.3 &      0.59   &      $-$0.23   &      25.217  &      24.386  &      0.832  &      1.077  &      0.562   \\
 V194 & $ab$ &  00 25 56.54 & $-$11 00 19.7 &      0.62   &      $-$0.21   &      25.292  &      24.418  &      0.874  &      0.648  &      0.345   \\
 V195 & $ab$ &  00 25 56.84 & $-$11 00 31.8 &      0.64   &      $-$0.19   &      25.258  &      24.360  &      0.898  &      0.741  &      0.317 
\enddata
\tablenotetext{a}{Approximate values are listed in italics.}
    % Table 9
% \end{deluxetable}
\end{deluxetable*}               % emulateapj

%%%
%%%   END Tables
%%%
%%%%%%%%%%%%%%%%%%%%%%%%%%%%%%%%%%%%%%%%%%%%%%%%%%%%%%%%%%%%%%%%%%%%%%%%%%%%%%%%
%%%%%%%%%%%%%%%%%%%%%%%%%%%%%%%%%%%%%%%%%%%%%%%%%%%%%%%%%%%%%%%%%%%%%%%%%%%%%%%%

 In the case of Cetus, the main factor for incompleteness is the spatial
 coverage, due to the off-center location of our ACS field to avoid a bright
 star and sample a radial region of the galaxy, and to the large extent of
 the galaxy. Figure~\ref{fig:6} shows the distribution of stars in Cetus
 from our ACS and WFPC2 fields on top of ellipses representing the core radius
 \citep[$r_c=1.3\arcmin \pm 0.1$,][]{mcc06} and integer multiples of the core
 radius. \citet{mcc06} estimated a tidal radius of $32.0\arcmin \pm 6.5$ and
 ellipticity $\varepsilon = 1-b/a=0.33 \pm 0.06$. Thus, the ACS field covers
 only about 1/300th of the area within the tidal radius.
 On the other hand, if we assume that the distribution of variable stars
 follows that of the RGB stars, we can estimate the total number of RR~Lyrae
 stars within the tidal radius as follows.
 First, we adopt the shape and orientation of the isodensity contours given by
 \citet{mcc05} to divide our sample of variable stars into three elliptical
 annuli containing approximately the same number of stars.
 To calculate the area of these annuli, we first distributed a large number of
 points ($10^7$) randomly thoughout the ACS field of view (FOV).
 Since the area of the FOV is known, the area of a given fraction of ellipse
 can be measured by counting the number of random points that it contains. We
 could then calculate the density profile of RR~Lyrae stars. Using this profile
 with the core and tidal radii from \citet{mcc05} as input to equation~(21) of
 \citet{kin62}, we calculate that our ACS sample represents about 17\% of the
 total number of RR~Lyrae stars ($\sim$1000).

 In the case of Tucana, the ACS field of view covers most of its area, so it is
 possible to measure its morphological parameters from our data.
 We first fitted ellipses to the stellar density maps. Even though the location
 of the center, position angle and ellipticity were left as free parameters, we
 found their values to be very stable, and decided to keep the average values
 as the best fit. We find that Tucana has an ellipticity $\varepsilon =
 1-b/a=0.48 \pm 0.01$, position angle of $-$3.$\degr \pm 1.$ and is centered on
 $\alpha_{J2000.0} = 22^h41^m49 \fs 92$,
 $\delta_{J2000.0} = -64 \arcdeg 25 \arcmin 14 \farcs 2$.
 Fig.~\ref{fig:7} shows the spatial distribution of stars. The variables and
 the core and tidal radii (see below) are also shown as larger dots and
 ellipses. We adopt the method of \citet{kin62} to determine the core and tidal
 radii from the spatial distribution of stars bright than the 80\% completeness
 limit, after correcting for completeness. This correction was obtained from
 the ratio of the number of recovered to injected stars in the artificial star
 test for each annular region. We find the best fit for $r_c=0.59\arcmin \pm
 0.01$ and $r_t=3.87\arcmin \pm 0.37$, which means that Tucana has a
 concentration factor similar to that of Fornax although it is 3-4 times
 smaller. The resulting King profile is presented in Fig.~\ref{fig:8}.
 Following the same method as discussed above, we calculate the density profile
 for the RR~Lyrae stars; it is shown as large open circles in Fig.~\ref{fig:8},
 shifted vertically to overlap the non-variable stars profile. This gives a
 total number of RR~Lyrae stars in Tucana of $\sim$400, implying that 10\% of
 the RR Lyrae variables are outside the field or in the gap between the chips.

\begin{figure*}%[b]   % Fig. 9
% \epsscale{0.8} % preprint
\epsscale{1.0} % emulateapj
\plotone{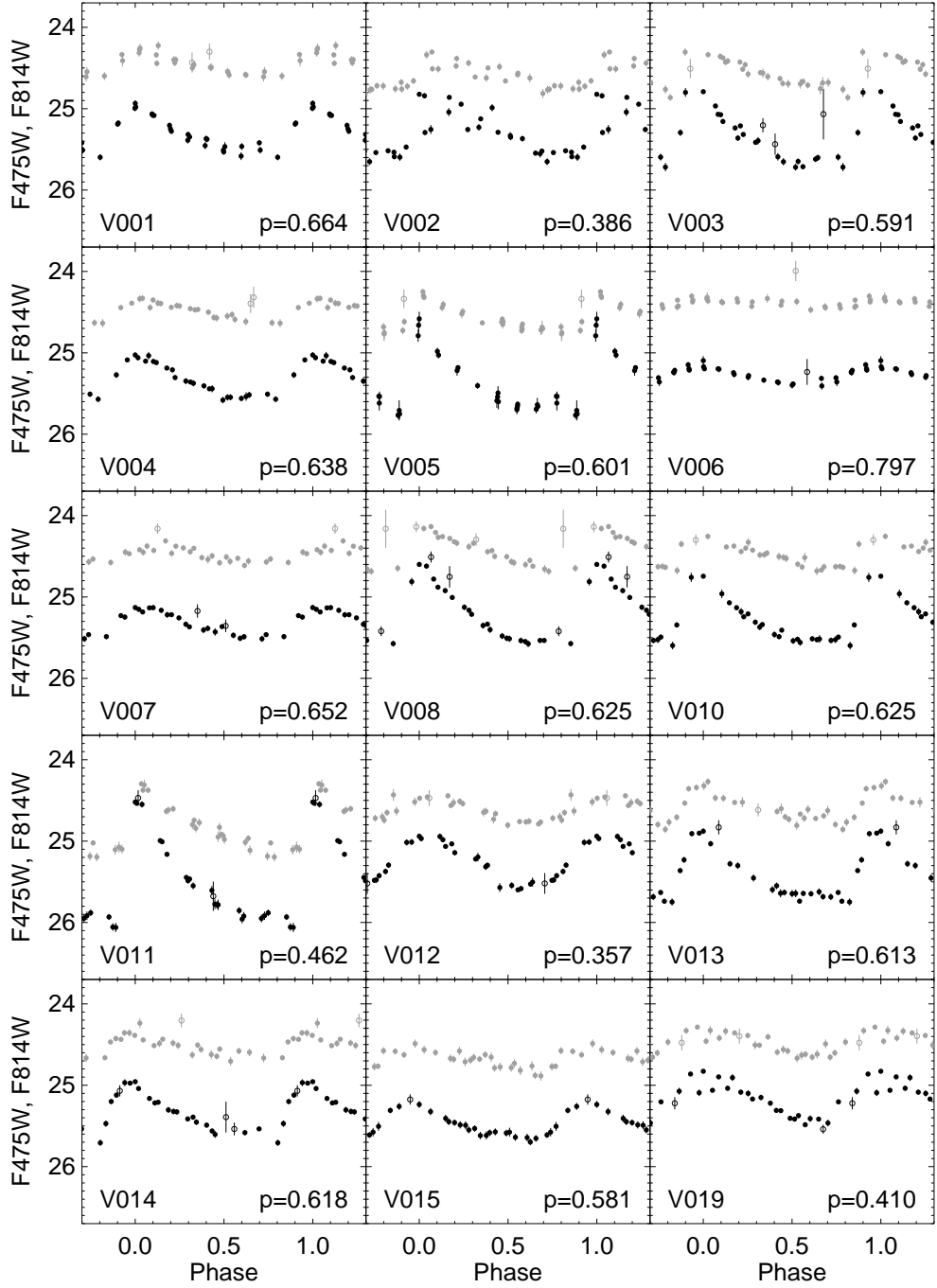}
\figcaption{Light-curves of the RR~Lyrae variables in the ACS field of Cetus in
 the F475W ({\it black}) and F814W ({\it grey}) bands, phased with the period
 in days shown in the lower right corner of each panel. Photometric error bars
 are shown. The open circles show the data points with errors larger than
 3-$\sigma$ above the mean error of a given star.
 {\it [Figures~\ref{fig:9}a--\ref{fig:9}l are available in the online version
 of the Journal]}.
\label{fig:9}}
\end{figure*}

\begin{figure*}%[b]   % Fig. 10
% \epsscale{0.8} % preprint
\epsscale{1.0} % emulateapj
\plotone{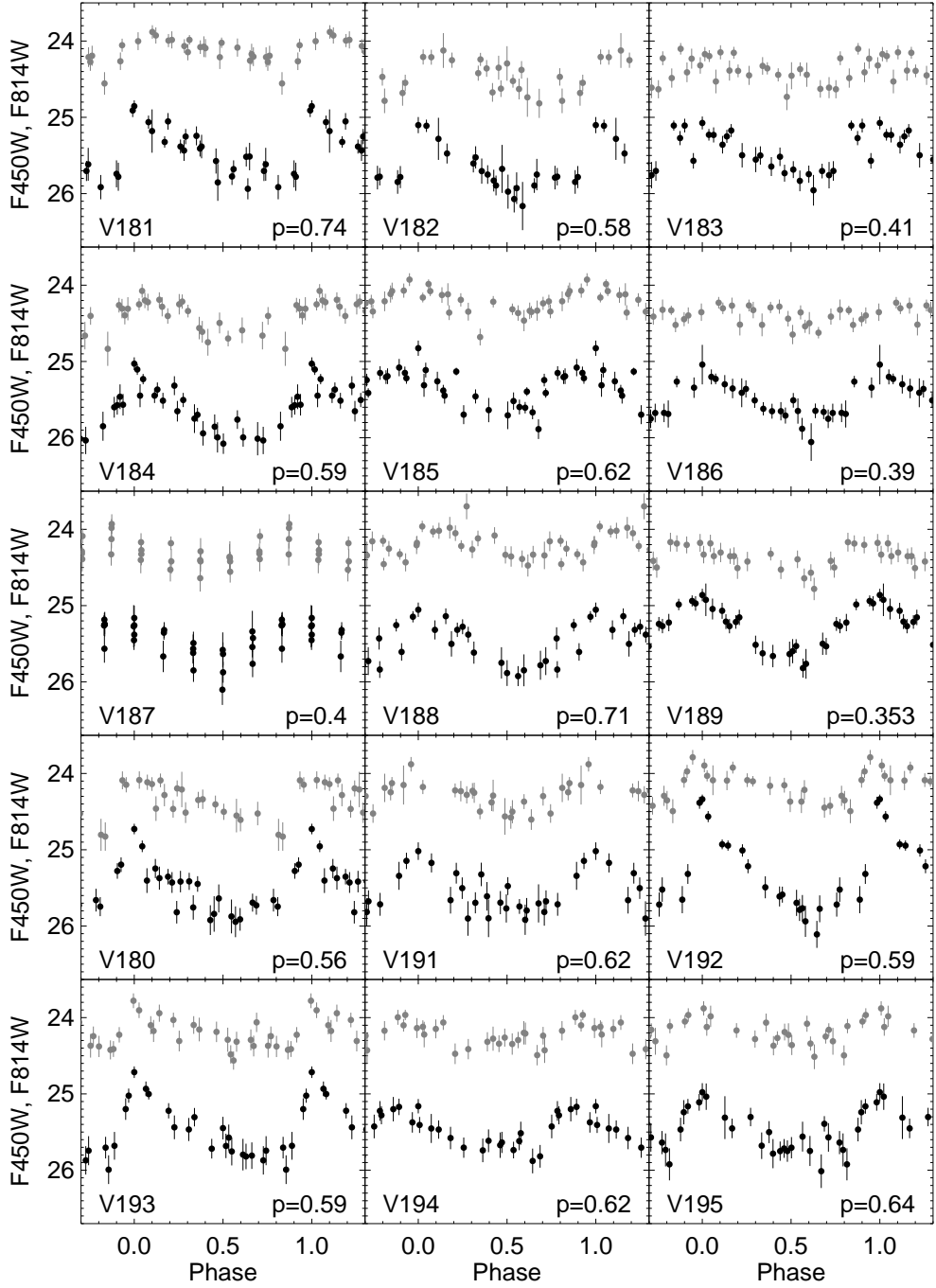}
\figcaption{Same as Fig.~\ref{fig:9}, but for all the variables of the
 outer field of Cetus from WFPC2 observations. For clarity, the F450W
 ({\it black}) and F814W ({\it grey}) light-curves have been shifted by 0.2 mag
 downward and upward, respectively, and the datapoints with large errorbars
 ($\sigma>$ 0.4 mag) omitted.
\label{sampleLC_cet2}}
\end{figure*}

\begin{figure*}%[b]   % Fig. 11
% \epsscale{0.8} % preprint
\epsscale{1.0} % emulateapj
\plotone{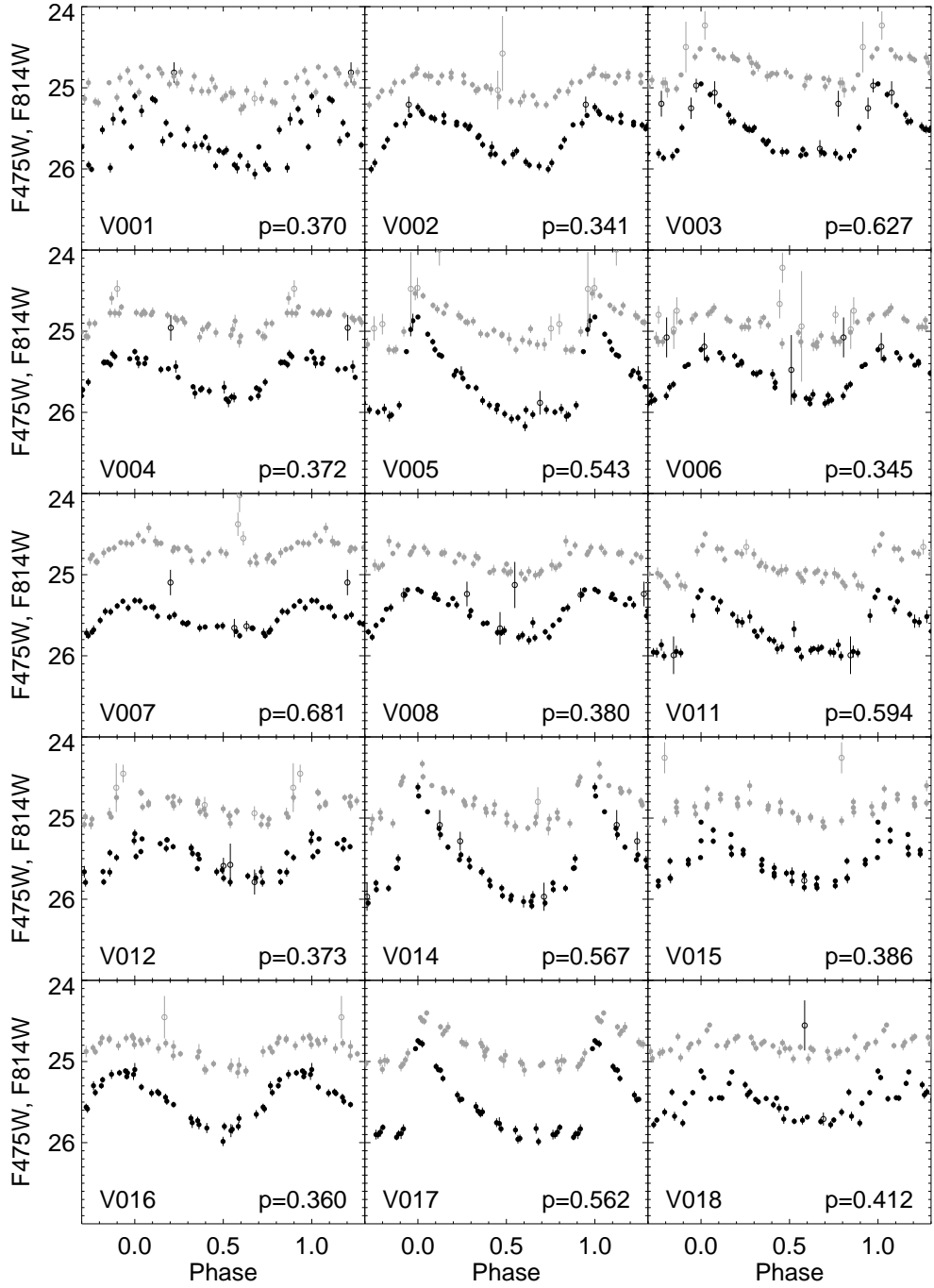}
\figcaption{Same as Fig.~\ref{fig:9}, but for Tucana.
 {\it [Figures~\ref{fig:11}a--\ref{fig:11}x are available in the online version
 of the Journal]}.
\label{fig:11}}
\end{figure*}

 To summarize, spatial coverage is the only factor really affecting the
 completeness of the sample of variable stars on and brighter than the HB,
 while crowding, signal-to-noise limitation, and temporal sampling prevented
 us from detecting fainter, shorter-period variables in both galaxies.

\section{RR Lyrae stars}\label{sec:5}

\subsection{Fundamental \& Overtone Pulsators}\label{sec:5.1}

 From the periods and light-curve shapes of the candidates, we identified 147
 RR Lyrae stars pulsating in the fundamental mode (RR$ab$), 8 in the
 first-overtone mode (RR$c$), and 17 in both modes simultaneously (RR$d$) in
 our ACS field of Cetus. The WFPC2 field contains 11 RR$ab$, 3 RR$c$, and 1
 RR$d$. For Tucana, the RR Lyrae stars were classified as 216 RR$ab$, 82 RR$c$,
 and 60 RR$d$. The light curves of all the RR~Lyrae stars are shown in
 Fig.~\ref{fig:9}--\ref{fig:11} for the ACS and WFPC2 samples of Cetus, and for
 Tucana, respectively. The RR$d$ are discussed in \S \ref{sec:5.2}.

 We note that a few of these RR~Lyrae variables are slightly brighter or
 fainter than the HB in each galaxy, and are labeled in Fig.~\ref{fig:12}.
 While the discrepancy can usually be linked to obvious problems in the
 photometry (e.g., V92 and V116 in Cetus, V125 in Tucana; see
 Appendix~\ref{sec:A}), for a few of these variables the shift in luminosity
 seems to be intrinsic.
 For example, V11 and V173 in Cetus are peculiar in the sense that they are
 $\sim$0.1 mag fainter and have much larger amplitude and shorter periods than
 the other RR$ab$ stars of this galaxy (see Fig.~\ref{fig:13} below), and may
 belong to a slightly more metal-rich population. V15 also appears fainter than
 the HB, although its period and amplitude are consistent with the majority of
 RR$ab$ stars.
 V55 in Cetus, and V341 and V364 in Tucana, on the other hand, appear a few
 tenths of magnitude above the HB.
 Their location on the period-luminosity and period-amplitude (PA) diagrams
 discard the possibility that they are anomalous or population II Cepheids, and
 indicate that they are bona fide RR Lyrae stars. Their amplitude and
 appearance on the stacked HST images also seem to discard blends. Therefore,
 these RR Lyrae are most likely evolved blue horizontal-branch (BHB) stars on
 their way to the AGB. Even though the evolutionary time of such stars within
 the IS is relatively short ($\la$10 Myr), the number of BHB stars and the
 extension of the HB to the blue make it a viable hypothesis. In the following
 we assume that the few outliers described above are bona fide RR~Lyrae stars.

\subsubsection{Cetus}

\begin{figure}%[b]   % Fig. 12
% \epsscale{0.6} % preprint
\epsscale{1.15} % emulateapj
\plotone{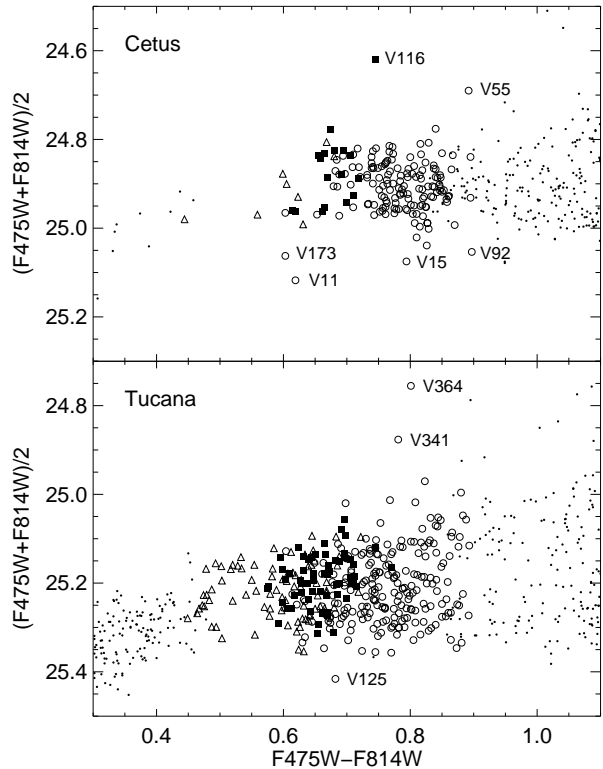}
\figcaption{Zoom-in on the HB of Cetus and Tucana, where the RR Lyrae variables
 have been overplotted. Open circles, open triangles, and filled squares
 represent RR$ab$, RR$c$, and RR$d$, respectively. A few outliers are labeled
 and discussed in  \S \ref{sec:5.1} and/or in Appendix~\ref{sec:A}.
\label{fig:12}}
\end{figure}

 In the case of Cetus, the mean periods for the RR$ab$ and RR$c$ are 0.614 and
 0.363 day, respectively. This value of $\langle P_{ab} \rangle$ is close to
 that expected based on the mean metallicity of the galaxy ([Fe/H]=$-$1.8;
 M. Monelli et al. 2009, in preparation) and the observed correlation between
 the mean period and the metallicity followed
 by the Galactic globular clusters (GGC) and the dSph of the LG (see
 Section~\ref{sec:5.3}). On the other hand, the fraction of RR$c$,
 calculated as $f_c = N_c / (N_{ab}+N_c)=0.05$, is surprisingly small. Typical
 values for this ratio range from 0.1 to 0.5 (see \S~\ref{sec:5.3}).
 Given the completeness at the magnitude of the HB discussed in Section
 \ref{sec:4}, it seems unlikely that many RR$c$ were undetected, even more so
 because fainter, lower amplitude RR$c$ were observed in Tucana. However, one
 can see on the CMD in Fig.~\ref{fig:12} that Cetus harbors a mainly red HB, at
 least in the region covered by the ACS field, having a HB ratio
 (HBR\footnotemark[14]) of $-$0.74. The hot side of the HB, which is where the
 RR$c$ are generally found, is therefore sparsely populated.
 In addition, taking into account the numerous RR$d$ yields
 $f_{cd} = N_{cd} / (N_{ab}+N_{cd})=0.15$, which is close to the typical value
 observed in the so-called Oosterhoff type I GCs \citep{oos39}.
 We recall that GGC are traditionally classified in one of the two Oosterhoff
 types according to the mean properties of their RR~Lyrae stars: Oosterhoff
 type I clusters tend to have intermediate metallicity ([Fe/H]$\sim-$1.5),
 RR~Lyrae with shorter periods ($\langle P_{ab} \rangle \sim$0.55, $\langle
 P_{c} \rangle \sim$0.32), and a low fraction of overtone pulsators
 ($f_c \sim$0.17), while Oosterhoff II clusters have low metallicity
 ([Fe/H]$\sim-$2.1), $\langle P_{ab} \rangle \sim$0.64, $\langle P_{c} \rangle
 \sim$0.37, and $f_c \sim$0.45 \citep{smi95}.
 The causes of this dichotomy is still a matter of debate \citep[see,
 e.g.,][]{cat05}.

 \footnotetext[14]{HBR = (B$-$R)/(B+V+R), where B, V, and R are the numbers of
 stars to the blue, within, and to the red of the IS \citep{lee90}.}

\subsubsection{Tucana}

\begin{figure}%[b]   % Fig. 13
% \epsscale{0.6} % preprint
\epsscale{1.1} % emulateapj
\plotone{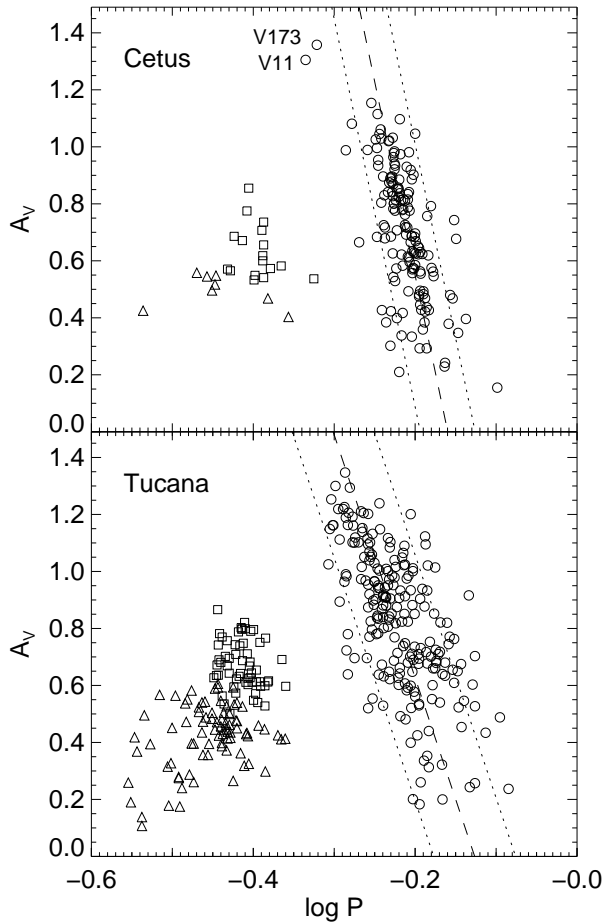}
\figcaption{Period-amplitude diagrams for the RR Lyrae stars of Cetus and
 Tucana. Circles, triangles and squares represent RR$ab$, RR$c$, and RR$d$
 (plotted with their overtone periods) respectively.
 The dashed line is a fit to the period-amplitude of the RR$ab$ after rejecting
 the points further than 1.5 sigmas ({\it dotted lines}), using period as the
 dependent variable.
\label{fig:13}}
\end{figure}

\begin{figure}%[b]   % Fig. 14
% \epsscale{0.6} % preprint
\epsscale{1.1} % emulateapj
\plotone{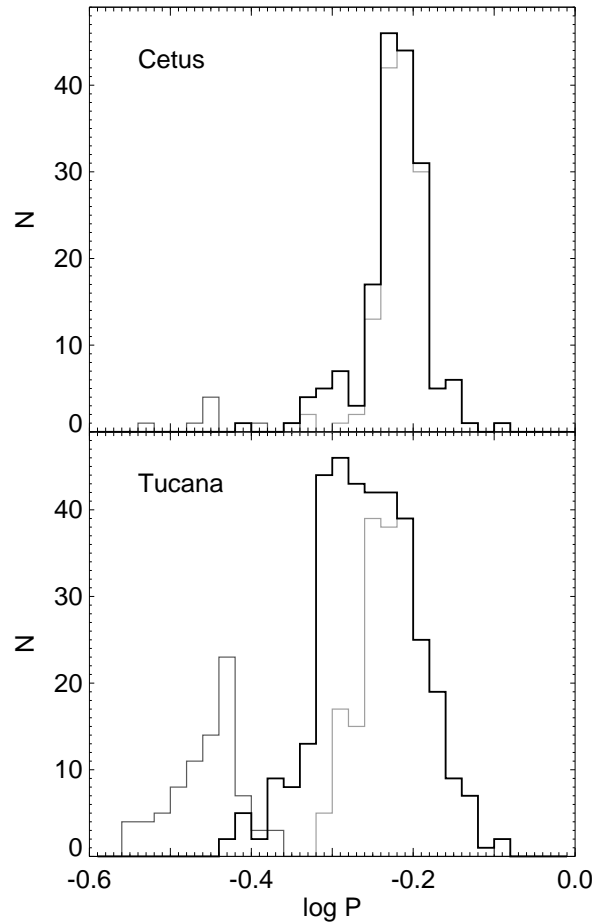}
\figcaption{Period histograms for the RR Lyrae stars of Cetus and Tucana.
 RR$ab$ and RR$c$ are shown as the light and dark gray histograms,
 respectively, while the black histograms represent the {\it fundamentalized}
 RR Lyrae stars (RR$ab$, RR$c$, and RR$d$).
\label{fig:14}}
\end{figure}

 For Tucana we calculated mean periods of 0.604 and 0.353 day for the RR$ab$
 and RR$c$, respectively.
 Contrary to Cetus, however, the ratio $f_c = 0.28$ is intermediate between the
 Oosterhoff types, and is a consequence of the well populated HB on both sides
 of the IS (see Fig.~\ref{fig:12}). Indeed, we find an average HBR of $-$0.14
 for the whole field-of-view, indicating that there is a comparable number of
 stars on each side of the IS. Including the RR$d$, we get a value close to
 that of the Oosterhoff type II GCs: $f_{cd} = 0.40$.
 Another point in which Tucana significantly differs from Cetus is the
 distribution of periods of the RR~Lyrae stars. Figure~\ref{fig:13} shows the
 PA diagram of the RR Lyrae stars in Cetus ({\it top}) and Tucana
 ({\it bottom}).
 The slope of the PA relation of the RR$ab$ is shallower and its dispersion
 much larger for the variables in Tucana than in Cetus.
 In \citetalias{ber08}, we showed that the large dispersion in the case of
 Tucana is partly due to the superposition of stellar populations with slightly
 different age and metallicities. Some of the scatter might also be introduced
 by the Blazhko effect \citep{bla07}, a modulation of the phase and amplitude
 of the pulsation, which can reduce the amplitude of the variations by up to
 half magnitude at a given period depending on the phase of the Blazhko cycles.
 However, it is unclear at the moment whether the difference in slope is real
 or a consequence of the larger dispersion.

 The period histograms are presented in Fig.~\ref{fig:14}. The solid histogram
 represents the {\it fundamentalized} RR Lyrae stars: the periods of the RR$c$
 were transformed to their fundamental mode equivalents by adding 0.128 to the
 logarithm of their periods, while the secondary (i.e., fundamental) periods
 were used for the RR$d$. The contributions of the RR$ab$ and RR$c$ are shown
 as the light and dark gray histograms, respectively.
 As expected from the relatively unpopulated RR$c$ region of the HB of Cetus,
 together with the steepness of its RR$ab$ PA relation, the period distribution
 is much tighter than that of Tucana.

\subsection{Double-Mode Pulsators}\label{sec:5.2}

\begin{figure*}%[b]   % Fig. 15
% \epsscale{0.9} % preprint
\epsscale{1.0} % emulateapj
\plotone{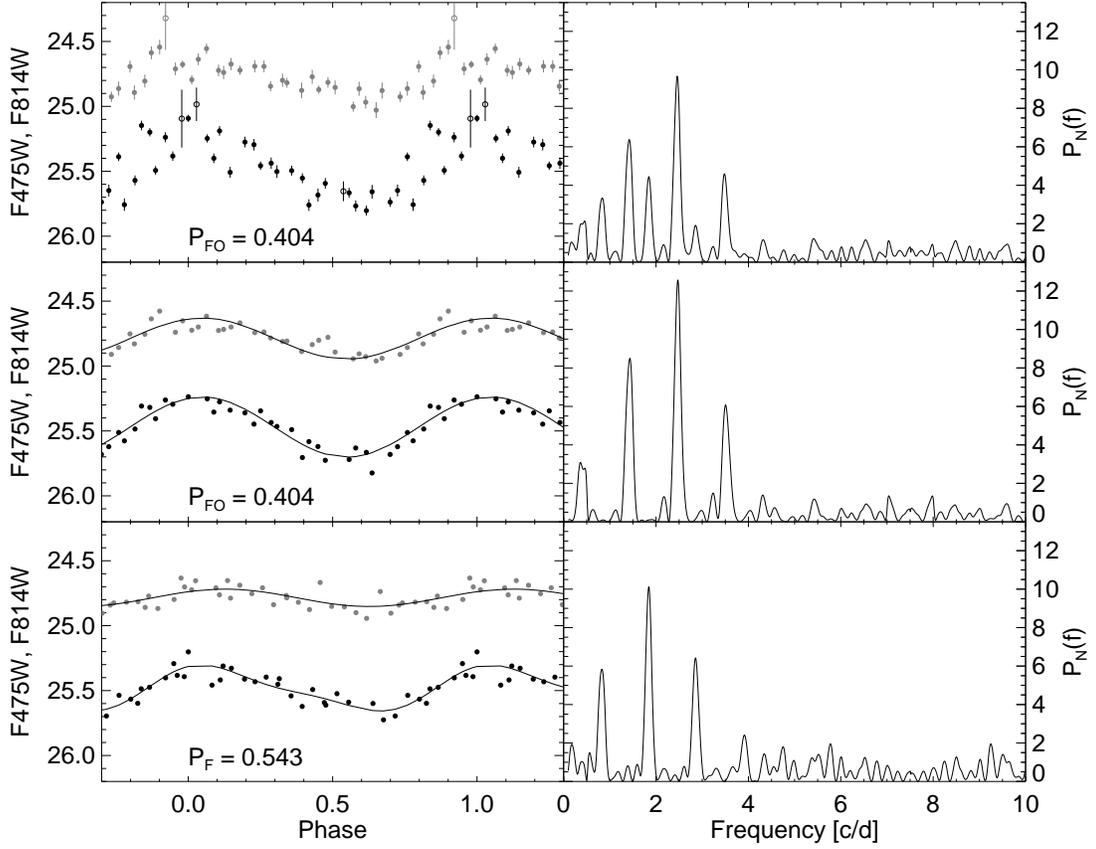}
\figcaption{{\it Left}: Light-curves of a RR$d$ in Tucana. Symbols are as in
 Fig.~\ref{fig:9}. The top panel shows the F475W and F814W data with no
 prewhitening and folded according to the primary period. The middle and
 bottom panel show the light-curves of the primary (secondary) period after
 prewhitening of the secondary (primary) period.
 {\it Right}: Periodograms corresponding to the F475W light curves of the left
 panels.
\label{fig:15}}
\end{figure*}

%%%%%%%%%%%%%%%%%%%%%%%%%%%%%%%%%%%%%%%%%%%%%%%%%%%%%%%%%%%%%%%%%%%%%%%%%%%%%%%%
%%%%%%%%%%%%%%%%%%%%%%%%%%%%%%%%%%%%%%%%%%%%%%%%%%%%%%%%%%%%%%%%%%%%%%%%%%%%%%%%
%%%
%%%   Table
%%%

% \begin{deluxetable}{lccccccccc}
\begin{deluxetable*}{lccccccccc} % emulateapj
\tabletypesize{\scriptsize}

\tablewidth{0pt}
\tablecaption{Properties of RR Lyrae Stars in Local Group Dwarf Spheroidals.\label{tab10}}
\tablehead{
\colhead{Galaxy} & \colhead{[Fe/H]} & \colhead{N$_{RRL}$\tablenotemark{a}} &
\colhead{$\langle P_{ab} \rangle$} & \colhead{log $\langle P_{ab} \rangle$} &
\colhead{f$_c$\tablenotemark{b}} & \colhead{\% RR$d$\tablenotemark{c}} &
\colhead{N$_{AC}$\tablenotemark{d}} &  \colhead{HBR\tablenotemark{e}} & \colhead{References}}
\startdata
Cetus             & $-$1.8  & 147+8+17  & 0.614 & $-$0.212 & 0.05  & 10 & 3  & $-$0.74 & 1        \\
Tucana            & $-$1.8  & 216+82+60 & 0.604 & $-$0.219 & 0.28  & 17 & 6  & $-$0.14 & 1        \\
\hline
Bo\"otes I        & $-$2.5  &   7+7+1   & 0.69  & $-$0.16  & 0.5   &  7 & 0  &  --     & 2,3      \\
Canes Venatici I  & $-$2.0  &  18+5     & 0.60  & $-$0.22  & 0.23  & -- & 3  &  --     & 4        \\
Canes Venatici II & $-$2.3  &   1+1     & 0.743 & $-$0.129 & 0.5   & -- & 0  &  --     & 5        \\
Carina            & $-$1.7  &  54+15+6  & 0.631 & $-$0.200 & 0.22  &  8 & 15 &  --     & 6        \\
Coma Berenices    & $-$2.53 &   1+1     & 0.670 & $-$0.174 & 0.5   & -- & 0  &  --     & 7        \\
Draco             & $-$2.0  & 214+30+26 & 0.615 & $-$0.211 & 0.12  & 10 & 9  &  --     & 8        \\
Fornax            & $-$1.3  & 396+119   & 0.585 & $-$0.233 & 0.231 & 20 & 17 &  --     & 9,10     \\
Leo I             & $-$1.82 &  47+7     & 0.602 & $-$0.220 & 0.13  & -- & 1  &  --     & 11       \\
Leo II            & $-$1.9  & 106+34+8: & 0.619 & $-$0.208 & 0.24  &  5 & 4  & $-$0.78 & 12,13    \\
Sculptor          & $-$1.8  & 132+74+18:& 0.585 & $-$0.233 & 0.40  &  8 & 3  &  0.06   & 14,15,16 \\
Sextans           & $-$1.6  &  26+7+3:  & 0.606 & $-$0.218 & 0.21  &  8 & 6  & $-$0.37 & 17,16    \\
U.Minor           & $-$2.2  &  47+35    & 0.638 & $-$0.195 & 0.43  & -- & 7  &  --     & 18
\enddata
\tablenotetext{a}{Number of RR$ab$ + RR$c$ + RR$d$ stars. `:' denotes
 uncertain values.}
\tablenotetext{b}{f$_c=$~N$_{c}$/(N$_{ab}$+N$_{c}$)}
\tablenotetext{c}{Approximate percentage of RR$d$.}
\tablenotetext{d}{Number of anomalous Cepheids.}
\tablenotetext{e}{Horizontal branch morphology: HBR $= (B-R) / (B+V+R)$,
 with B, V, R representing the numbers of HB stars to the blue, within,
 and to the red of the instability strip \citep{lee90}.}
\tablerefs{
(1) This work;
(2) \citealt{dal06};
(3) \citealt{sie06};
(4) \citealt{kue08};
(5) \citealt{gre08};
(6) \citealt{dal03};
(7) \citealt{mus09};
(8) \citealt{kin08};
(9) \citealt{ber02};
(10) \citealt{cle06};
(11) \citealt{hel01};
(12) \citealt{sie00};
(13) \citealt{mig96};
(14) \citealt{kal95};
(15) \citealt{kov01};
(16) \citealt{har01};
(17) \citealt{mat95};
(18) \citealt{nem88}.}
    % Table 10
% \end{deluxetable}
\end{deluxetable*}               % emulateapj

%%%
%%%   END Table
%%%
%%%%%%%%%%%%%%%%%%%%%%%%%%%%%%%%%%%%%%%%%%%%%%%%%%%%%%%%%%%%%%%%%%%%%%%%%%%%%%%%
%%%%%%%%%%%%%%%%%%%%%%%%%%%%%%%%%%%%%%%%%%%%%%%%%%%%%%%%%%%%%%%%%%%%%%%%%%%%%%%%

 Both Cetus and Tucana were found to possess a significant number of variable
 HB stars having a dispersion in magnitude, once phased with the best period,
 much larger than the one expected from photometric errors alone. In addition,
 these variables are located in the central part of the IS where RR$ab$ and
 RR$c$ overlap (see Fig.~\ref{fig:12}).
 The occurence of the Blazhko effect was discarded as the main contribution
 since the observations were collected in a time range spanning five days at
 maximum, while the period of the Blazhko modulations ranges typically from
 tens to hundreds of days.
 In addition, their periodograms show the characteristic double-peaks of RR$d$
 variables (see Fig.~\ref{fig:15}, top right panel).

\begin{figure}%[b]   % Fig. 16
% \epsscale{0.6} % preprint
\epsscale{1.1} % emulateapj
\plotone{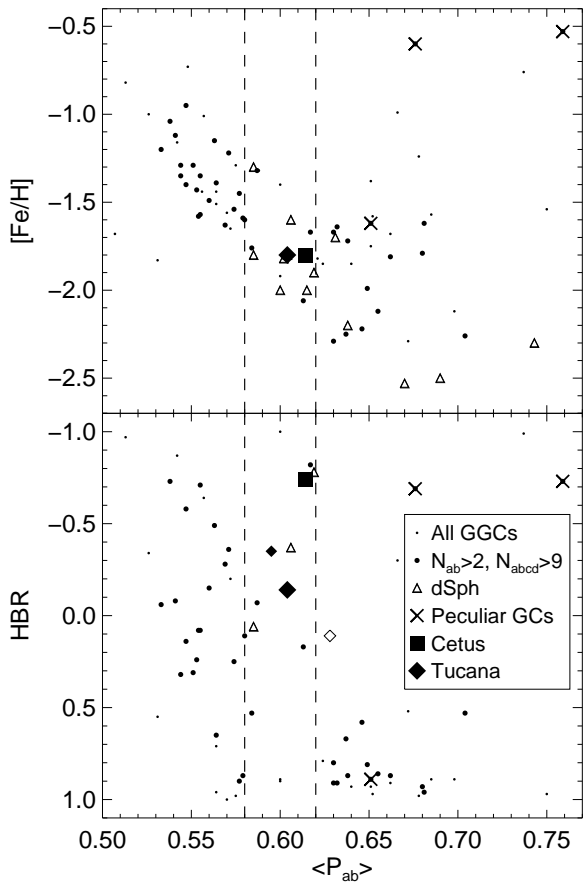}
\figcaption{Mean RR$ab$ periods--metallicity ({\it top}) and mean RR$ab$
 periods--HB morphology ({\it bottom}) diagrams in Cetus and Tucana.
 The peculiar GGCs $\omega$\,Cen, NGC\,6388 and NGC\,6441 are indicated by
 crosses. The dashed lines delimit the Oosterhoff Gap. The smaller diamonds
 in the bottom panel are for the inner ({\it filled}) and outer ({\it open})
 annuli of Tucana, from \citetalias{ber08}.
\label{fig:16}}
\end{figure}

\begin{figure}%[b]   % Fig. 17
% \epsscale{0.6} % preprint
\epsscale{1.1} % emulateapj
\plotone{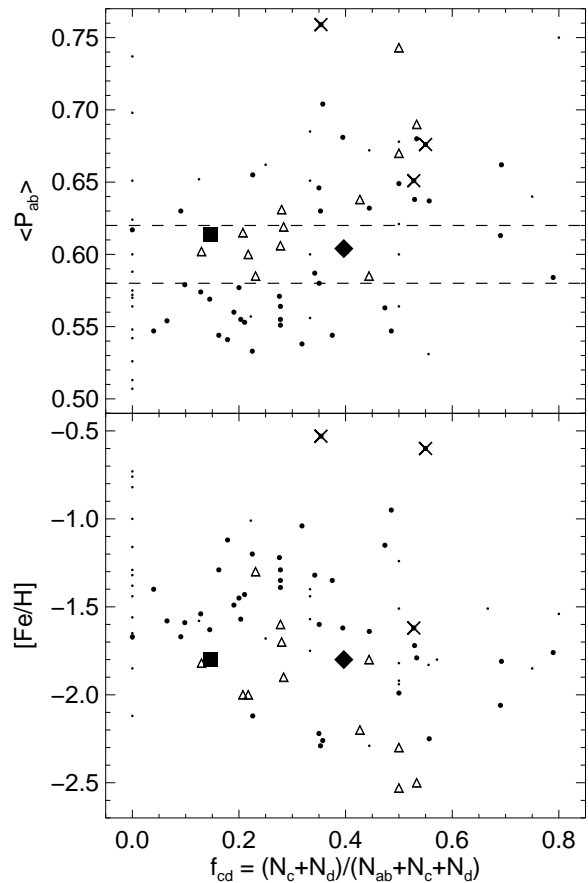}
\figcaption{Fraction of RR$c$+RR$d$ versus mean RR$ab$ periods and metallicity
 in Cetus and Tucana. Symbols are as in Fig.~\ref{fig:16}. The dashed lines in
 the top panel delimit the Oosterhoff Gap.
\label{fig:17}}
\end{figure}

 The periods of each mode were searched using an iterative process, first
 prewhitening the primary period to find the secondary period, then refining
 the primary period after prewhitening of the secondary period.
 In the left panels of Fig.~\ref{fig:15} we present the F475W \& F814W
 light-curves of a RR$d$ in Tucana: the top panel shows the light-curves
 without prewhitening, while the middle and bottom panels show the light-curves
 of the primary (secondary) period after prewhitening of the secondary
 (primary) period. The curves show the low-order Fourier series fit that were
 subtracted from the data to find the period of the other component.
 The periodograms corresponding to each F475W light-curve are shown in the
 right panels.

 Because of the short timebase of our observations and the rather small number
 of datapoints, it was not possible to obtain very accurate periods for each
 pulsation mode. However, the ratio of the primary-to-secondary periods was
 found to be consistent with the typical range of known RR$d$
 ($0.74 < P_1/P_0 < 0.75$; see, e.g., Fig. 13 of \citet{cle04}.

 While the fraction of RR$d$ in Cetus is similar to that observed in other dSph
 (see Table~\ref{tab10}), it is about twice as high in Tucana. Such a large
 fraction of double-mode pulsators has only been observed in one dSph
 \citep[Fornax, $\sim$20\%;][]{cle06} and in a few GGC ($\sim$29\% in M 68:
 \citealt{wal94}; $\sim$18\% in IC 4499: \citealt{wal96}; 15--18\% in M 15:
 \citealt{cor08}).
 No consensus has been reached yet concerning the origin of the relative
 abundance of RR$d$ in stellar systems, but the narrowness of the possible
 metallicity range \citep[0.0002--0.001;][]{pop00}, the mass range at a
 given metal content, and the temperature range \citep[$\sim$0.02~M$_\sun$ and
 $\Delta$T$_{eff}\la$~60~K;][]{sza04} are possibly the main suspects. The
 fulfillment of these three conditions at the same time is probably a transient
 phenomenon, explaining the scarcity of stellar system with a large fraction of
 double-mode pulsators.

\subsection{Comparison with LG dSph \& GGC}\label{sec:5.3}

 In Fig.~\ref{fig:16} and \ref{fig:17}, we compare the average properties of
 the RR~Lyrae stars in Cetus and Tucana with those of other LG dSph and GGC.
 The data for the GGC ({\it small and large dots}) come from the compilation of
 \citet{cle01}, updated with values from the literature when the change in
 mean period, HBR, or number of variables was significant (M\,2:
 \citealt{lee99}, \citealt{laz06}; M\,75: \citealt{cor03}; NGC\,2808:
 \citealt{cor04}; NGC\,6388 and NGC\,6441: \citealt{cor06}). However, given the
 lack of theoretical \citep[e.g.,][]{kov98,bon97a} and observational
 \citep{kal04} evidence supporting the existence of second-overtone RR~Lyrae
 stars (``RR$e$"), variables identified as such in their list were considered
 RR$c$ here. The open triangles represent values for the LG dSphs from the
 literature, which are summarized in Table~\ref{tab10}.

 Figure~\ref{fig:16} shows the metallicity and HB morphology as a function of
 the mean period of the fundamental mode RR Lyrae stars. The peculiar GGCs
 NGC\,6388, NGC\,6441, and $\omega$\,Cen are indicated by crosses.
 Note how the location of Bo\"otes~I, at $\langle P_{ab} \rangle$=0.69
 \citep{dal06} and [Fe/H]=$-$2.5 \citep{mun06} seems to indicate that the
 correlation between [Fe/H] and $\langle P_{ab} \rangle$ of the Oosterhoff
 type I GGC \citep[e.g.,][]{cle01} actually extends to the Oosterhoff II
 domain.
 Cetus and Tucana are shown as a filled square and a filled diamond,
 respectively. While they both follow the trend in P$_{ab}$--[Fe/H] defined by
 the GGC, the bottom panel shows that the HB of Cetus is too red given the mean
 period of its RR$ab$.
 The only GGC with a sufficient number of RR~Lyrae stars in this part of the
 P$_{ab}$--HBR space is the peculiar GGC Rup~106, which is generally considered
 to be $\sim$2~Gyr younger than the other GGC \citep[e.g.,][]{sal97}.

 On the other hand, the mean period of the RR$ab$ in Tucana is close to that
 expected from its HB morphology. The values for the inner and outer annuli of
 Tucana---excluding the intermediate region for clarity---from
 \citetalias{ber08}, shown as smaller diamonds, are also in rough agreement
 with the GGCs and LG dwarfs.
 Interestingly, it seems that the HB morphologies of the dSphs tend to be
 redder than the HBs of GGCs at a given period. Given that the HB gets redder
 for younger ages at constant metallicity, it is possible that the
 difference between the GGCs and the dSphs is due to the generally more
 extended star formation histories of the latter.

 Following \citet{pet03}, in the bottom panel of Fig.~\ref{fig:17} we plotted
 the fraction of overtone RR Lyrae stars versus the metallicity for the same
 dSph and GGC as in Fig.~\ref{fig:16}. While \citet{pet03} used only RR$ab$
 and RR$c$, we found that including the RR$d$ (when present) with the RR$c$
 gave a slightly tighter correlation. We estimated the significance of this
 correlation using the Pearson product-moment correlation coefficient $\rho$,
 which is obtained by dividing the covariance of the two variables by the
 product of their standard deviations. We found that the correlation between
 the fraction of overtone pulsators and [Fe/H] of the clusters with a
 well-sampled population of RR~Lyrae stars---excluding the six clusters with
 [Fe/H]$> -1.0$ or f$_{cd}>0.6$---increases from $-$0.23 to $-$0.29 when the
 double-mode pulsators are included, at a significance greater than 99.5\%.
 As shown in the bottom panel of Fig.~\ref{fig:17}, f$_{cd}$ tends to be larger
 at lower metallicities. Since clusters with lower metallicity generally have a
 bluer HB, the hotter region of the IS tends to be more populated, therefore
 containing more RR$c$ stars.
 Both Cetus and Tucana agree with the trend, even though the former has very
 few RR$c$ (f$_c\sim$~0.05) and the latter has a unusually large number of
 RR$d$ (17\%).

 Interestingly, it seems that $\langle P_{ab} \rangle$ and f$_{cd}$ are also
 correlated, as shown in the top panel of Fig.~\ref{fig:17}. The value of the
 Pearson correlation coefficient for these two quantities is $\rho$=0.53 at a
 significance level of 99.99\% when including the dSph. This correlation
 comes from the temperature distribution of the stars along the HB:
 a bluer (i.e., hotter) HB contains a larger fraction of overtone pulsators,
 while the pulsational period decreases with increasing temperature.

\begin{figure}%[b]   % Fig. 18
% \epsscale{0.8} % preprint
\epsscale{1.2} % emulateapj
\plotone{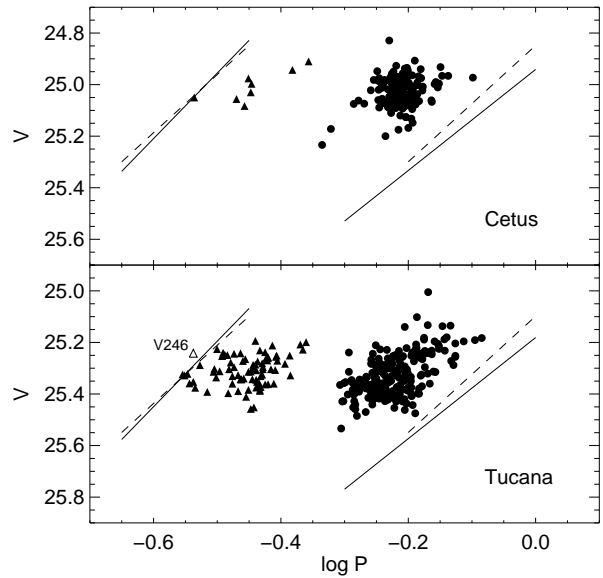}
\figcaption{Distribution of RR Lyrae in the $\langle V \rangle$-log P plane,
 with the predicted edges of the instability strip of \citet[{\it dashed
 lines}]{cap00} and \citet[{\it solid lines}]{dic04}.
 {\it See text for details.}
\label{fig:18}}
\end{figure}

\section{Distance Estimates}\label{sec:6}

 As stated above, the photometric and pulsational properties of RR~Lyrae stars
 are fundamental tools to estimate distances.
 In this section, we used the two main methods adopted in the literature to
 calculate the distance: the luminosity-metallicity relation, which arises
 from the knowledge that the intrinsic luminosity of HB stars mainly depends
 on their metallicity, and the period-luminosity-metallicity (PLM) relation,
 based on the theoretical location of the IS in the period-luminosity plane. In
 both cases, we used the intensity-averaged mean magnitude in the Johnson V
 band calculated in \S \ref{sec:2}. For this reason, RR$d$ variables were not
 taken into account.

\subsection{The Luminosity-Metallicity Relation}\label{sec:6.1}

 The luminosity-metallicity relation generally has the form
 M$_V$ = a + b[Fe/H], where $a$ and $b$ assume different values depending on
 the calibration \citep[see][for a review of the various calibrations]{san06}.
 The most recent calibrations are not linear but present a break around
 [Fe/H]$\sim$$-$1.5, although this is not of concern here since the mean
 metallicity of both galaxies is below this value.
 In the present paper, we use the slope given by \citet{cle03}, who derived it
 from new photometry and spectroscopy of a large number ($>$100) of RR~Lyrae
 stars close to the LMC bar.
 The zero-point was chosen such that the LMC distance modulus is
 (m-M)$_{LMC,0}$ = 18.515$\pm$0.085 (assuming [Fe/H]$_{LMC}$=$-$1.5), which
 corresponds to the weighted mean of a large number of independent and reliable
 distance estimates to the LMC \citep[see][]{cle03}:
\begin{equation}
 M_V = 0.866(\pm0.085) + 0.214(\pm0.047) [Fe/H].
\end{equation}
 This calibration gives M$_V$ = 0.545 at [Fe/H]=$-$1.5, in very good agreement
 with calibrations of the RR~Lyrae absolute magnitude as a function of
 metallicity using the absolute magnitude of SX~Phoenicis stars and main
 sequence fitting using Hipparcos trigonometric parallax data
 \citep[see][for a review]{tam08}.

\begin{figure}%[b]   % Fig. 19
% \epsscale{0.9} % preprint
\epsscale{1.2} % emulateapj
\plotone{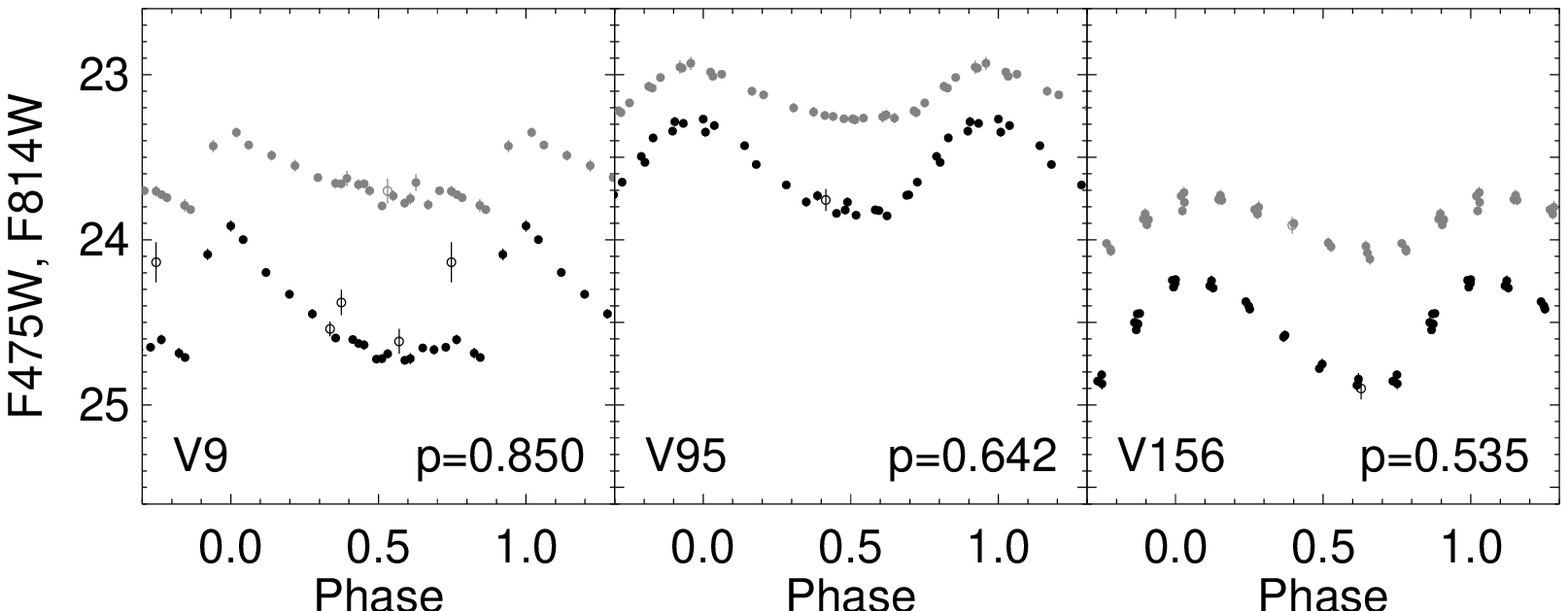}
\plotone{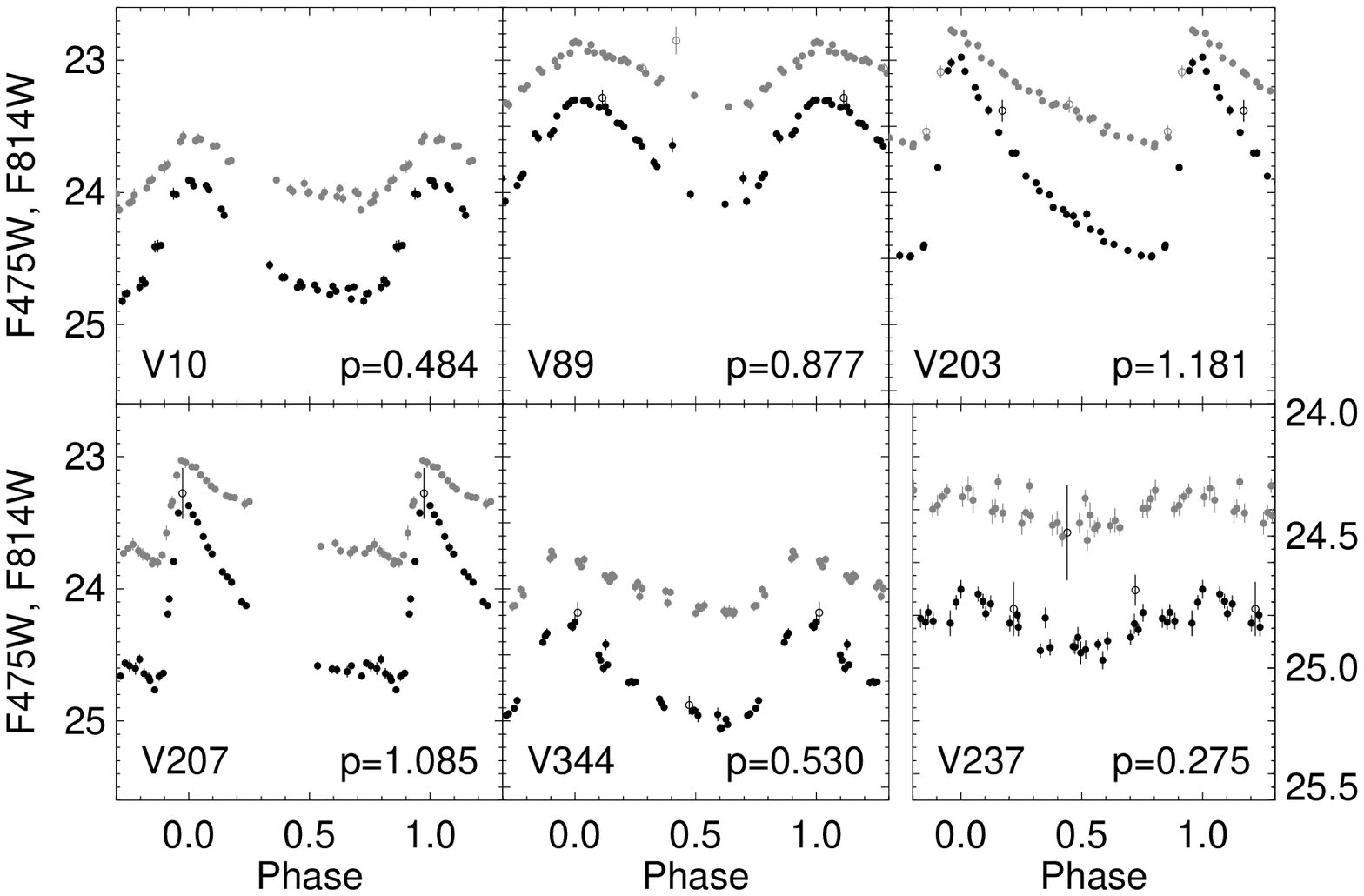}
\figcaption{Above-horizontal branch stars in Cetus ({\it Top}) and Tucana
 ({\it Bottom}). Symbols are as in Fig.~\ref{fig:9}. Note the different
 vertical scale for V237.
\label{fig:19}}
\end{figure}

\begin{figure}%[b]   % Fig. 20
% \epsscale{0.7} % preprint
\epsscale{1.3} % emulateapj
\plotone{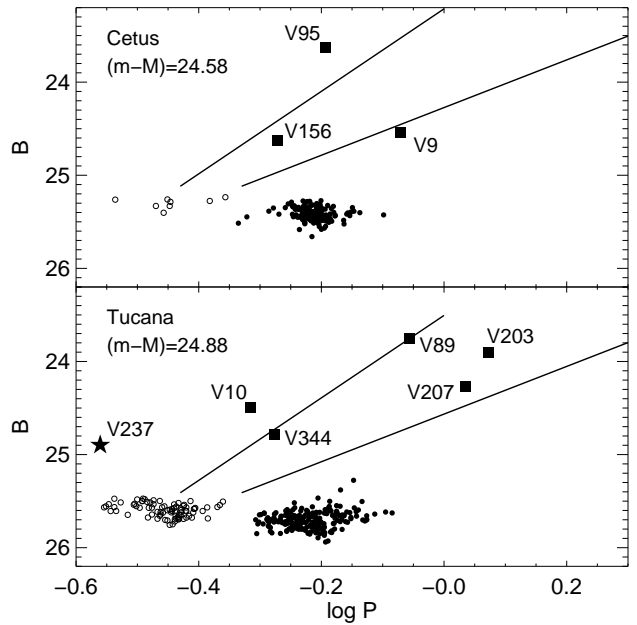}
\figcaption{Period-magnitude diagram for Cetus and Tucana showing the location
 of the anomalous Cepheids ({\it filled squares}) and V237 ({\it star}). Note
 that the magnitudes are in the standard Johnson system. The lines from
 \citep{bon97c} show the loci of the fundamental and first-overtone anomalous
 Cepheids. Open and filled circles show the RR$c$ and RR$ab$, respectively.
\label{fig:20}}
\end{figure}

 For Cetus and Tucana, we calculated the mean magnitude of the RR~Lyrae stars
 to be $\langle V \rangle$=25.028$\pm$0.005 and $\langle V
 \rangle$=25.321$\pm$0.005, respectively. Given the large number of variables
 in both cases, the outliers (see Fig.~\ref{fig:12}) have very little influence
 on these values: we iteratively rejected stars further than 3- then
 1-$\sigma$, and found that the mean magnitudes changed by less than 0.01. As
 this is within the uncertainties, all the RR$ab$ and RR$c$ were used in the
 following.

 The adopted mean metallicity was obtained from our star formation histories
 (M. Monelli et al. 2009, in preparation). For the old population of both Cetus
 and Tucana, we found Z=0.0003$\pm$0.0001 (i.e., [Fe/H]=$-$1.82$\pm$0.15
 assuming Z$_\sun$=0.0198 \citep{gre93} and
 [Fe/H]=log Z+1.70 $-$ log(0.638 f + 0.362) \citep{sal93} with
 the $\alpha-$enhancement factor f set to zero). This gives a HB luminosity of
 M$_V$=0.48$\pm$0.12, where the uncertainty was quantified with Monte Carlo
 simulations and takes into account the uncertainty on the zero-point and slope
 of eq.~(1) and on the metallicity of the galaxies. This gives distance moduli
 of 24.55$\pm$0.12 for Cetus and 24.84$\pm$0.12 for Tucana.

 Interestingly, when separately calculating the distance to the bright and
 faint subsamples of RR~Lyrae stars in Tucana presented in \citetalias{ber08},
 we obtain very similar values.
 Assuming they have slightly different average metallicities (Z=0.0002 and
 0.0005), we find that the distance modulus of the brighter
 ($\langle V \rangle$=25.251$\pm$0.005), more metal-poor subsample differs from
 that of the fainter ($\langle V \rangle$=25.379$\pm$0.003), more metal-rich
 subsample by only 0.05 (24.81 vs. 24.86, respectively), which is consistent
 within the uncertainty.
 This strengthens the claims of \citetalias{ber08} that the difference in
 luminosity of the two subsamples is due to a difference in metallicity.

\subsection{The Period-Luminosity-Metallicity Relation}\label{sec:6.2}

 We also determined the distance modulus of each galaxy by matching the PLM
 relation for evolutionary pulsators at the first-overtone blue edge (FOBE) of
 the IS from \citet[their eq. 3]{cap00}. The theoretical limits of the IS are
 shifted in magnitude until the FOBE coincides with the observed distribution
 of first-overtone RR~Lyrae stars.
 Figure~\ref{fig:18} shows the position of the IS ({\it dashed lines})
 overplotted on the distribution of observed RR~Lyrae stars in the $\langle V
 \rangle$--log~P plane.

 Adopting the mean metallicity given above, we derive (m$-$M)$_{Cet}$=24.54 and
 (m$-$M)$_{Tuc}$=24.79, for which \citet{cap00} give an intrinsic dispersion of
 $\sigma _V$=0.07 mag due to uncertainties associated with the various
 ingredients of the model.
 Combined with the errors in metallicity, mean magnitude and period, we
 estimate a total uncertainty in the distances of $\sim$0.1.

 At this metallicity, we can also use a RR$c$ mass of 0.7~M$_\sun$
 \citep{bon97b} as input to the updated PLM from \citet[their eq.~3 and
 4]{dic04}. Assuming a mixing-length parameter $l/H_p$=1.5, the location of the
 predicted edges of the IS is shown with the solid lines in Fig.~\ref{fig:18}
 and basically yields the same values for the distance moduli (24.53 and
 24.77).

 Note that we excluded V246 in Tucana, since the low amplitude
 (A$_{475}\sim$0.15) and poor quality of the light-curves (see
 Fig.~\ref{fig:11}) prevented an accurate determination of its period;
 assuming the period and mean magnitude were correctly measured, the distance
 should be shortened by $\sim$0.1.
 On the other hand, the distance determination in Cetus is based on only one
 star (V111). While statistical considerations hamper the reliability of this
 distance calculation (which strictly speaking might be considered an
 upperlimit), the location of this star at the very edge of the blue side of
 the IS (see Fig.~\ref{fig:12}), together with the high quality of its
 light-curves, strengthen the relevance of the derived distance. Indeed, both
 the luminosity-metallicity and PLM relations give very similar values for the
 distance to Cetus.

 Interestingly, for both galaxies the temperature of the fundamental-mode red
 edge (FRE) of the IS seems to be underestimated in both calibrations of the
 PLM. Because of the large uncertainty in the efficiency of convection in the
 external layers of stars, represented by the mixing-length parameter $l/H_p$
 in stellar evolution models, and the higher sensitivity of the red side of
 the IS to $l/H_p$, \citet{cap00} tentatively place the red edge of the
 pulsation region at $\delta$log P=0.45 with respect to the FOBE.
 However, the low metallicity GCs ([Fe/H]$\la-$1.3, their Fig.~1) indicate
 that the $\delta$log P is actually closer to $\sim$0.4 at the metallicity of
 our galaxies.
 Therefore, using the FRE at $\delta$log P=0.45 from the FOBE gives a distance
 modulus smaller by $\sim$0.1 than the distance calculated from the FOBE
 itself, as was already observed by \citet{dal03}.
 Similarly, \citet{dic04} note that for some of the clusters of their
 sample---which also happen to be the most metal-poor clusters
 ([Fe/H]$\la-$1.5)---the value of the distance modulus to fit the observed
 distribution of ab-type variables is smaller by $\sim$0.15 mag than the value
 derived from the FOBE under the assumption of a constant mixing-length
 parameter \citep[see][for a discussion of the effect of the mixing-length
 parameter on the distance estimates]{dic04}.

\begin{figure}%[b]   % Fig. 21
% \epsscale{0.8} % preprint
\epsscale{1.2} % emulateapj
\plotone{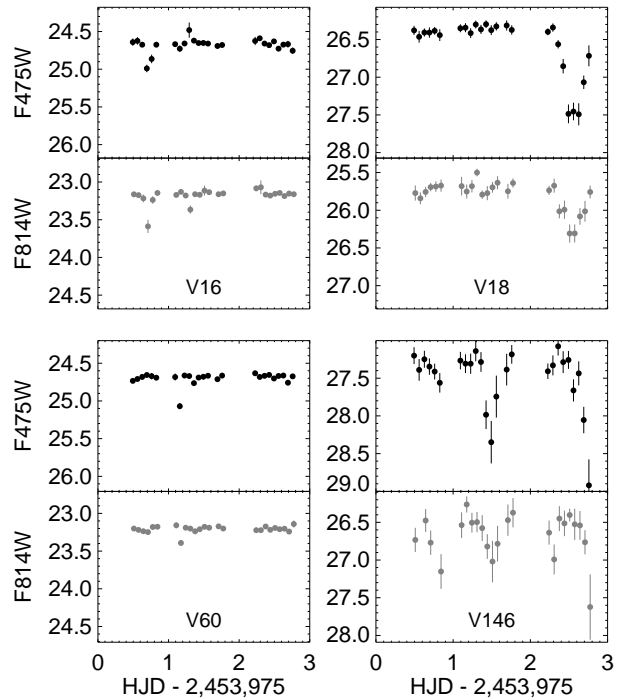}
\figcaption{Cetus candidate binary stars. The vertical scale is the same for
 all the panels.
\label{fig:21}}
\end{figure}

\begin{figure*}%[b]   % Fig. 22
% \epsscale{1.0} % preprint
\epsscale{1.0} % emulateapj
\plotone{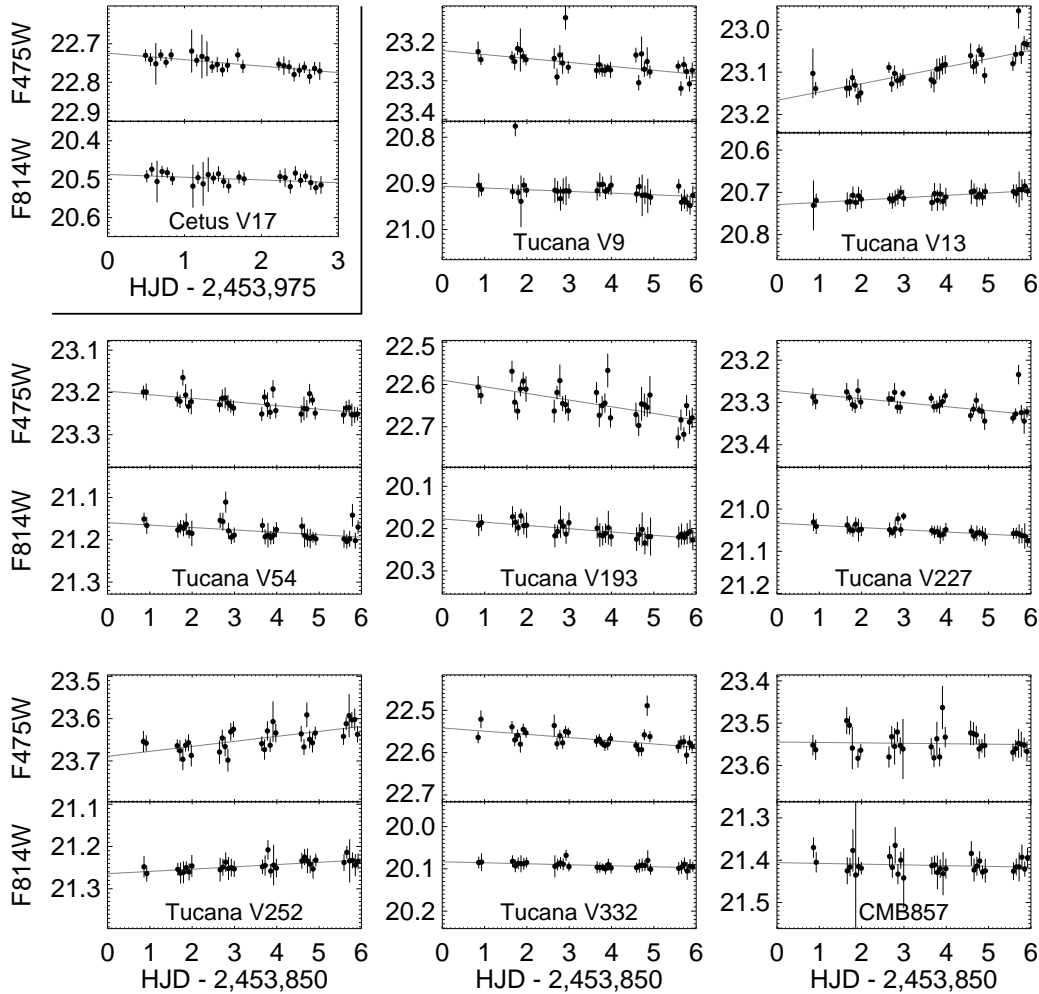}
\figcaption{LPV candidates in Cetus and Tucana. The last light-curve is the
 Mira candidate of \citet{cas96}, for which we do not observe any variability.
\label{fig:22}}
\end{figure*}

\subsection{Results}\label{sec:6.3}

 In paragraphs \ref{sec:6.1} and \ref{sec:6.2} we derived the
 distance modulus of Cetus and Tucana adopting two independent methods, based
 on empirical and theoretical calibrations, which gave very similar results.
 Given the statistical considerations presented above, we retain the results of
 the luminosity-metallicity method as our best distances.

 According to \citet{sch98}, the extinction along the line of sight of Cetus
 and Tucana in V is 0.088 and 0.097, respectively. In the following, we adopt
 reddening-corrected distance moduli of 24.46$\pm$0.12 and 24.74$\pm$0.12,
 respectively, corresponding to 780$\pm$40~kpc and 890$\pm$50~kpc.
 These are in good agreement with previous determinations based on
 independent methods (e.g., TRGB; (m$-$M)$_{0,Cet}$=24.39$\pm$0.07:
 \citealt{mcc05}; (m$-$M)$_{0,Tuc}$=24.69$\pm$0.16: \citealt{sav96}).

\section{Above-HB variables}\label{sec:7}

 Cetus and Tucana respectively harbor three and six variables located above
 the HB. The different light-curve shapes, shown in Fig.~\ref{fig:19}, hint
 that they probably belong to different types of variables or pulsate in
 different modes.
 In Fig.~\ref{fig:20} we plot their intensity-averaged Johnson B magnitudes
 (see Section~\ref{sec:2}) versus the logarithm of their period. The solid
 lines show the loci of the fundamental and first-overtone mode anomalous
 Cepheids (AC) from \citet{bon97c}.

 From their position in this figure, we tentatively classify V9 in Cetus,
 and V203 and V207 in Tucana as fundamental mode ACs, while the remaining
 AHB---except V237 in Tucana---are closer to the first-overtone ACs locus.
 It also seems that no type II Cepheid, which have longer periods for a given
 magnitude \citep[e.g.,][]{nem88}, is present in these galaxies. This is not
 unexpected, since Fornax and Draco are the only dSph in which they have been
 observed to date \citep{ber02,har98}.

 V237 in Tucana has a very low amplitude ($\sim$0.15 in F475W) and its
 luminosity is only 0.6 magnitude brighter than the HB. While it clearly
 presents a variability and looks isolated on our images, no period could be
 found that gave a smooth light-curve.
 Even though its position in the CMD makes it a good candidate low-amplitude
 Cepheid, there is still the possibility that it actually is a foreground
 variable with a very short period or multiple periods.

 The observed number of AC can be compared with that expected from the visual
 luminosity of each galaxy through the specific frequency, calculated as the
 number of AC per $10^5L_{V,\sun}$ \citep{mat95}.
 Assuming V237 in Tucana is an AC, and correcting for completeness (10\% in
 Cetus and 90\% in Tucana, see Section~\ref{sec:4}), we find the specific
 frequencies to be $\sim$5 and $\sim$1, respectively.
 From the updated plot of \citet{pri05} and the absolute visual
 magnitude of our two galaxies (M$_{V,Cet}=-$11.3$\pm$0.3: \citealt{mcc05};
 M$_{V,Tuc}=-$9.6$\pm$0.3: \citealt{sav96}), we find that Tucana
 falls very close to the fit, while Cetus has more AC than expected.
 This might be an indication that the tidal radius of Cetus from \citet{mcc06}
 was overestimated, leading to an overestimation of the correction for
 completeness in Section~\ref{sec:4}. In this respect, preliminary analysis of
 deep (V$\sim$26) VIMOS images of Cetus indicates that the field located
 $\sim$10$\arcmin$ from the center contains very few, if any, stars belonging
 to Cetus (E. Bernard et al. 2009, in preparation).

\section{Other Candidate Variables}\label{sec:8}

 In addition to the classical IS variables, we also detected four candidate
 eclipsing binary systems and one LPV in Cetus, and seven candidate LPV in
 Tucana. Figure~\ref{fig:21} shows the light-curves of the eclipsing binaries.
 Note that V16 and V60 might not be binary stars as they are not located on or
 near the main-sequence as most binary stars. In addition, a minimum lasting
 about three hours seems unlikely for a star on the RGB. We decided to not
 discard them as minimum light occurs in both bands at the same time, and the
 individual images corresponding to these minimums do not present chip defects
 or cosmic rays at the location of the candidates.
 V18 and V146, on the other hand, are definitely eclipsing binaries, although
 we could not determine their orbital period as we did not observe a well
 defined secondary minimum.

 Among the bright red stars, one in Cetus and seven in Tucana presented a
 larger variability index. All are located at or slightly above the TRGB. Their
 light curves are presented in Fig.~\ref{fig:22}. Even though the amplitude of
 these variations is really small over the length of the observing run
 ($\la 0.1$ magnitude), the fact that the amplitude varies in the same sense in
 both bands and by a larger amount in F475W supports their classification as
 LPV. None of the LPV variables of Tucana correspond to the ones discovered by
 \citet{cas96}. Of their three candidates, one is saturated on our images and
 another falls inside the gap between the two chips. The third (\#857 in their
 catalog) does not seem to present variability in our data. For reference, its
 light-curve from our photometry is also shown at the bottom of
 Fig.~\ref{fig:22}. Assuming it is a bona fide Mira variable, it is possible
 that we observed it at minimum light, which can last for several weeks at
 almost constant magnitude.
 Another possibility is that their candidate variable is actually a flickering
 RGB star, which present variability over 10-minute time scales \citep{mig04}.
 Our 20-minute long exposures would erase such variations, while these could
 appear on the shorter exposures of \citet{cas96}.

\section{Discussion and Conclusions}\label{sec:9}

 Since the dSph were first discovered, our perceptions of them have changed
 from simple, GC-like objects to complex systems presenting a wide variety of
 properties. Cetus and Tucana are no exceptions, even though they appear
 similar at first sight: their metal content is comparable, and \citet{whi99}
 noted that the closest match to the RGB of Cetus among the dSph is the RGB of
 Tucana. %In addition, note the
%  similarity of their main-sequences, revealed for the first time on the LCID
%  data (see Fig.\ref{fig:1}).

 However, both their HB morphology and RR~Lyrae populations expose differences
 in a more subtle level between these two galaxies. As discussed above, the HB
 of Cetus is redder than expected from its metallicity, especially compared to
 the HB of Tucana. While the red side of the HB is well-populated in both
 galaxies, Cetus has very few stars on the blue side although they extend as
 far to the blue as in Tucana.
 Given that in a dSph the main parameter affecting the HB are age and
 metallicity, in the sense that the HB gets bluer for older ages and lower
 metallicities, this might be a hint that the first burst of star formation
 started at the same time in both galaxies, albeit with a much higher intensity
 in Tucana. The same conclusion is reached from the SFH analysis of both
 galaxies (M. Monelli et al. 2009, in preparation).
 This strong early burst might also explain why the overall metallicity of
 Tucana is so similar to that of Cetus even though it is much less massive, and
 the presence of multiple old populations in Tucana \citepalias{har01,ber08}.

 The mean properties of their RR Lyrae are also very different, as evidenced by
 Figs.~\ref{fig:16} and \ref{fig:17}. The main disparities are the mean
 {\it fundamentalized} periods (0.601 and 0.555 for Cetus and Tucana,
 respectively) and the fraction of overtone pulsators, both consequences of the
 temperature (i.e., mass) distribution on the HB.

 In addition, both galaxies present internal variations of the HB morphology
 as a function of galactocentric distance. We discussed the gradients in the
 RR~Lyrae properties of Tucana in a previous paper \citepalias{ber08}. In the
 case of Cetus, the small spatial coverage did not allow an accurate study of
 these gradients. However, because the ACS field lies slightly outside the
 center of the galaxy, we could use the division in elliptical annuli described
 in Section~\ref{sec:4} to calculate the HBR and mean period as a function
 of radius. We find that the HBR does increase slightly with radius
 ([$-$0.82$\pm$0.05,$-$0.75$\pm$0.04,$-$0.63$\pm$0.04], from the inner- to the
 outermost annulus).
 On the other hand, $\langle$P$_{ab}\rangle$ is constant within the errors
 throughout the sampled radius. Observations of a larger area are necessary to
 check if the change in HBR is real and accompanied by a change in mean period.
 Thus, to date Tucana is the only dSph in which a radial gradient in the mean
 period of its variables has been observed.

 Cetus is located about 780 kpc from the MW, and \citet{mcc06} calculate that
 it is also 680 kpc from M31, while Tucana is 890 kpc from the MW and on
 the opposite side of the MW from M31. Thus, the spatial isolation of both
 galaxies inside the LG probably sheltered them from the strong interactions
 consequence of close encounters with massive galaxies like the MW or
 M31---tidal stripping and stirring, ram-pressure stripping \citep{may06}.
 Nevertheless, it is possible that the two dwarfs are on very radial orbits, as
 seen for at least 10\% of subhaloes in cosmological simulations \citep{ghi98},
 or that they were on more tightly bound orbits but were scattered out due to
 three-body interactions \citep{sal07}.
 In the latter cases, they would have suffered tidal interactions or ram
 pressure stripping at some point in their evolution, having experienced at
 least one close passage with the MW, but surely not as much as the bulk of the
 dSphs population of MW and M31.
 Until a good understanding of their orbits is available it is safe to assume
 that their differing properties, together with similarities between
 satellites and isolated galaxies, might actually indicate that environmental
 effects must be complemented by some other basic mechanism(s) affecting
 their individual evolution.

\acknowledgments

{\it Facility:} \facility{HST (ACS, WFPC2)}

 The authors are grateful to the anonymous referee for an extensive report that
 helped improve this manuscript, and to Antonio Sollima for useful comments.
 Support for this work was provided by a Marie Curie Early Stage Research
 Training Fellowship of the European Community's Sixth Framework Programme
 under contract number MEST-CT-2004-504604, the IAC (grant 310394), the
 Education and Science Ministry of Spain (grants AYA2004-06343 and
 AYA2007-3E3507), and NASA through grant GO-10505 from the Space Telescope
 Science Institute, which is operated by AURA, Inc., under NASA contract
 NAS5-26555.\
 This research has made use of the NASA/IPAC Infrared Science Archive, which
 is operated by the Jet Propulsion Laboratory, California Institute of
 Technology, under contract with the National Aeronautics and Space
 Administration.
 This research used the facilities of the Canadian Astronomy Data Centre
 operated by the National Research Council of Canada with the support of the
 Canadian Space Agency.

\appendix

\section{Comments on Individual Variables}\label{sec:A}

 The following comments are based on the careful inspection of the stacked
 images, of the light curves, and/or peculiar properties exhibited on one or
 more of the plots presented in this work.

\subsection{In Cetus}

V1 --- Possible blend with a faint star.

V4 --- Close to a bright star.

V6 --- Low Amplitude. Looks isolated on the images.

V11 --- Peculiar: large amplitude and $\sim$0.1 mag below the HB.

V32 --- Low Amplitude. Looks isolated on the images.

V35 --- Blend.

V39 --- In bad column. Very noisy light-curve.

V40 --- Blend.

V45 --- Blend.

V47 --- Blended with V48.

V48 --- Blended with V47.

V53 --- Possible blend with faint stars.

V55 --- $\sim$0.2 mag above the HB. Looks isolated on the images.

V71 --- Low Amplitude. Looks isolated on the images.

V92 --- Located close to the edge of chip 1. Some points bad or missing
        because of dithering.

V93 --- Located close to the edge of chip 1. Some points bad or missing
        because of dithering.

V105 --- Blend.

V116 --- $\sim$0.3 mag above the HB. Possible blend with faint stars.

V130 --- Possible blend with faint stars.

V133 --- Located close to the edge of chip 1. Some points bad or missing
         because of dithering.

V135 --- Located close to the edge of chip 1. Some points bad or missing
         because of dithering.

V154 --- Blend. Low amplitude.

V158 --- Blend.

V159 --- Blend.

V165 --- Possible blend with faint stars.

V173 --- Peculiar: large amplitude and $\sim$0.1 mag below the HB.

V176 --- Blend. Low amplitude.

\subsection{In Tucana}

V25 --- Blend.

V28 --- Located close to the edge of chip 2.

V36 --- Possible blend with a faint star.

V38 --- Close to a bright star.

V50 --- Close to a bright star.

V51 --- In bad column. Very noisy light-curve.

V56 --- Possible blend. Noisy light-curve and low amplitude.

V58 --- Possible blend with faint stars.

V61 --- Blend.

V65 --- Close to bad column.

V67 --- Low amplitude. Looks isolated on the images.

V69 --- Blend.

V70 --- Blend. Might not be a RRd.

V72 --- Located close to the edge of chip 2. Some points bad or missing
        because of dithering.

V78 --- Possible blend with faint stars, close to bad column.

V80 --- In bad column.

V81 --- In bad column.

V82 --- Close to a bright background galaxy. RRd.

V92 --- Blend.

V96 --- Blend.

V99 --- Possible blend with a faint star.

V110 --- Noisy light-curves and low amplitude. Looks isolated on the images.

V111 --- Possible blend with faint stars.

V112 --- Noisy light-curves. Possible blend with a faint star.

V113 --- Blend.

V114 --- Possible blend with faint stars.

V118 --- Blend. Noisy light-curve in F814W.

V119 --- Close to a bright star. RRd.

V120 --- Blend.

V121 --- Located close to the edge of chip 2.

V124 --- Possible blend with faint stars.

V125 --- Rising phase + peak missing. Inaccurate mean mag and amplitudes
         because of bad fits.

V126 --- Close to a bright background galaxy.

V129 --- Possible blend with faint stars.

V137 --- Located close to the edge of chip 1. Some points bad or missing
         because of dithering.

V139 --- Blend.

V145 --- Close to a bright star.

V158 --- Blend. RRd.

V160 --- Close to a background galaxy.

V167 --- Possible blend with faint stars.

V168 --- Blend.

V177 --- Blend.

V179 --- Blend.

V180 --- Blend.

V184 --- Close to a bright star.

V185 --- Close to a bright star.

V187 --- Blend.

V189 --- Blended with a background galaxy.

V194 --- Blend.

V196 --- Possible blend.

V199 --- Blend. RRd.

V201 --- Blend.

V220 --- Possible blend with faint stars.

V227 --- Blend. LPV.

V231 --- Blend.

V232 --- Blend.

V233 --- Possible blend with a faint background galaxy.

V238 --- Blend. Noisy light-curves.

V239 --- Blend. RRd.

V241 --- Possible blend with faint stars.

V246 --- Low amplitude. Possible blend with faint stars. Noisy light-curves.

V248 --- Blend. Noisy light-curve in F814W.

V251 --- Blend.

V253 --- Blend.

V264 --- Blend.

V267 --- Blend.

V284 --- Blend.

V285 --- Possible blend with faint stars.

V289 --- Possible blend with faint stars.

V293 --- Blend. RRd.

V295 --- Possible blend with faint stars.

V304 --- Blend.

V306 --- Blend.

V326 --- Blend.

V334 --- Possible blend with faint stars.

V341 --- $\sim$0.3 mag above the HB. Looks isolated on the images.

V353 --- Blended with a background galaxy.

V364 --- $\sim$0.45 mag above the HB. Looks isolated on the images.

\section{Finding Charts}

 The finding charts for the whole sample of variable stars are presented
 in the electronic version of the Astrophysical Journal (Figs.~23 \& 24).

\begin{figure*}%[b]   % Fig. 23a
\epsscale{1.0} % preprint
%\epsscale{1.16} % emulateapj
\plotone{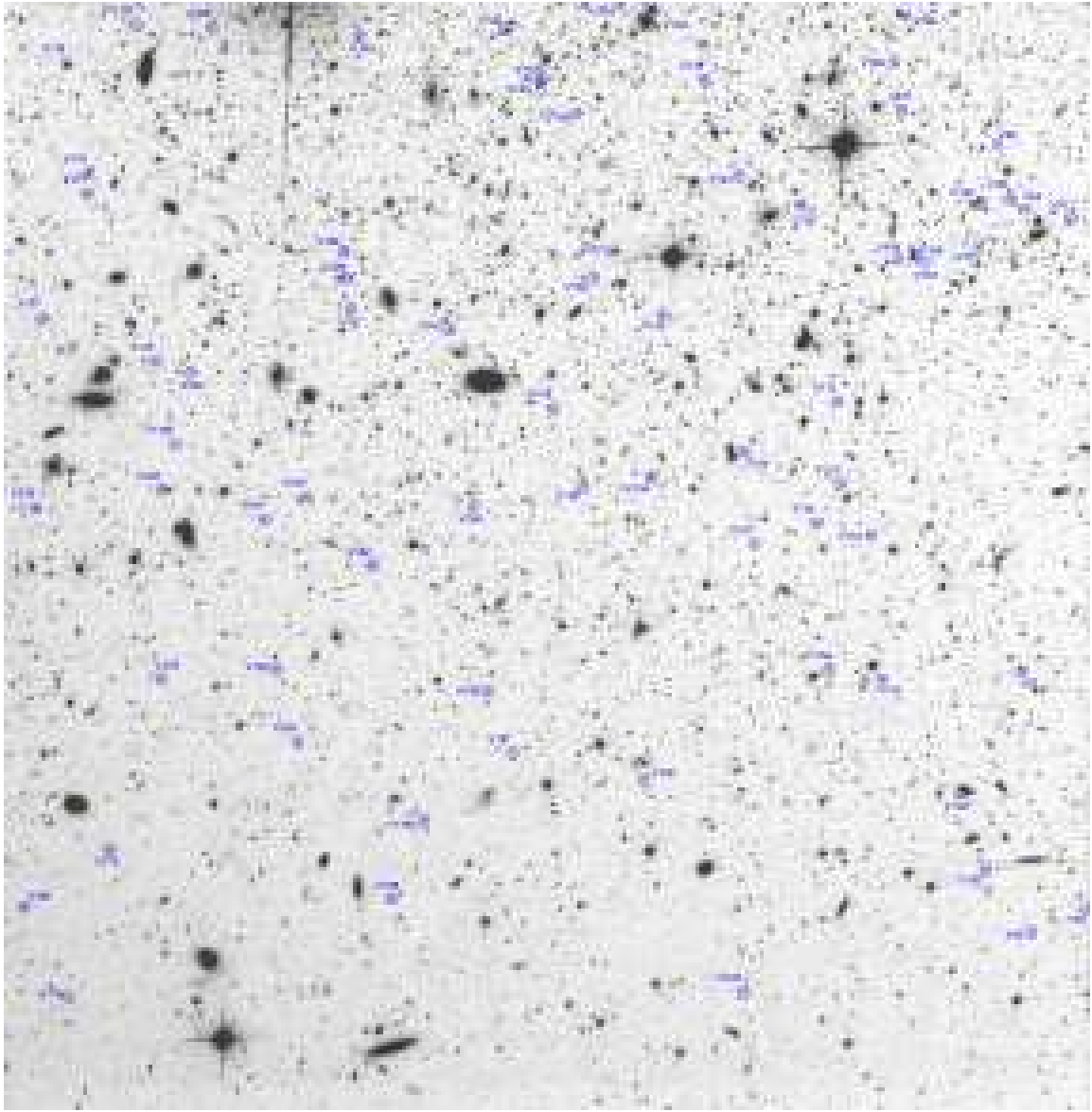}
\figcaption{Finding chart for the North-East quadrant in Cetus. North is up and
 East to the left. {\it [Figures~\ref{fig:23a}a--\ref{fig:23a}d are available in
 the online version of the Journal]}.
\label{fig:23a}}
\end{figure*}

% \begin{figure*}%[b]   % Fig. 23b
% \epsscale{1.0} % preprint
% %\epsscale{1.16} % emulateapj
% \plotone{f23b.eps}
% \figcaption{Finding chart for the South-East quadrant in Cetus. North is up and
%  East to the left [On-line only].
% \label{fig:23b}}
% \end{figure*}
% 
% \begin{figure*}%[b]   % Fig. 23c
% \epsscale{1.0} % preprint
% %\epsscale{1.16} % emulateapj
% \plotone{f23c.eps}
% \figcaption{Finding chart for the North-West quadrant in Cetus. North is up and
%  East to the left [On-line only].
% \label{fig:23c}}
% \end{figure*}
% 
% \begin{figure*}%[b]   % Fig. 23d
% \epsscale{1.0} % preprint
% %\epsscale{1.16} % emulateapj
% \plotone{f23d.eps}
% \figcaption{Finding chart for the South-West quadrant in Cetus. North is up and
%  East to the left [On-line only].
% \label{fig:23d}}
% \end{figure*}

\begin{figure*}%[b]   % Fig. 24a
\epsscale{1.0} % preprint
%\epsscale{1.16} % emulateapj
\plotone{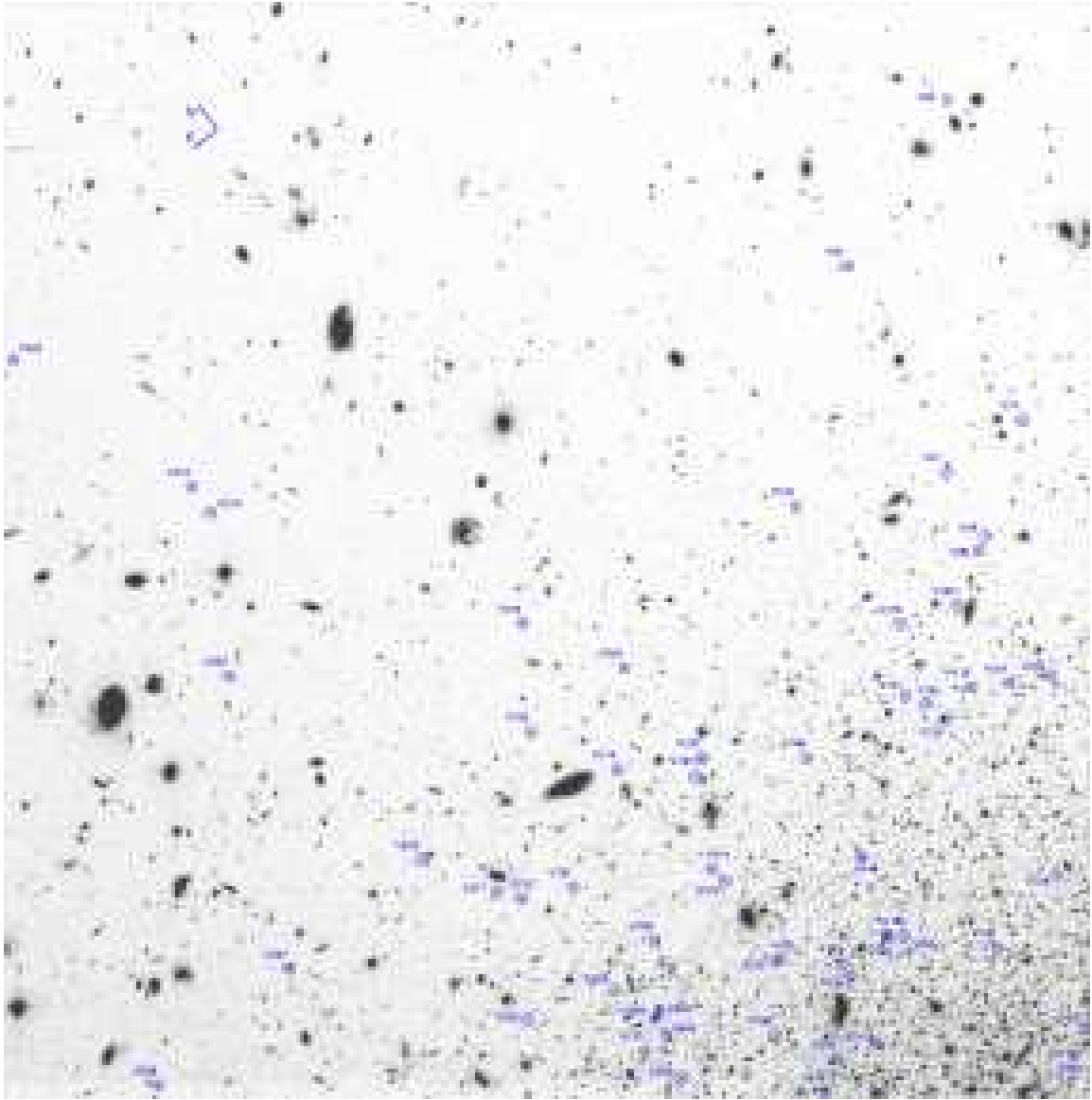}
\figcaption{Finding chart for the North quadrant in Tucana.
 {\it [Figures~\ref{fig:24a}a--\ref{fig:24a}d are available in the online version
 of the Journal]}.
\label{fig:24a}}
\end{figure*}

% \begin{figure*}%[b]   % Fig. 24b
% \epsscale{1.0} % preprint
% %\epsscale{1.16} % emulateapj
% \plotone{f24b.eps}
% \figcaption{Finding chart for the East quadrant in Tucana [On-line only].
% \label{fig:24b}}
% \end{figure*}
% 
% \begin{figure*}%[b]   % Fig. 24c
% \epsscale{1.0} % preprint
% %\epsscale{1.16} % emulateapj
% \plotone{f24c.eps}
% \figcaption{Finding chart for the West quadrant in Tucana [On-line only].
% \label{fig:24c}}
% \end{figure*}
% 
% \begin{figure*}%[b]   % Fig. 24d
% \epsscale{1.0} % preprint
% %\epsscale{1.16} % emulateapj
% \plotone{f24d.eps}
% \figcaption{Finding chart for the South quadrant in Tucana [On-line only].
% \label{fig:24d}}
% \end{figure*}

\end{document}